\documentclass[11pt,a4paper]{article}

\usepackage{jcappub}
\usepackage[english]{babel}
\usepackage{subcaption}
\usepackage{float}
\usepackage[export]{adjustbox}
\usepackage[normalem]{ulem} 
\usepackage{cancel}

\usepackage{xcolor}
\definecolor{mycolor}{RGB}{255,0,255}

\newcommand{\mb}[1]{\mathbf{#1}} 
\newcommand{\mr}[1]{\mathrm{#1}} 
\newcommand{\mc}[1]{\mathcal{#1}} 
\newcommand{\vg}[1]{\boldsymbol{#1}} 

\newcommand{\f}[2]{\frac{#1}{#2}} 

\newcommand{\bp}[1]{\left({#1}\right)} 
\newcommand{\bs}[1]{\left[{#1}\right]} 
\newcommand{\ba}[1]{\left<{#1}\right>} 
\newcommand{\bc}[1]{\left\{{#1}\right\}} 
\newcommand{\bv}[1]{\left|{#1}\right|} 

\AtBeginDocument{\renewcommand{\*}{\cdot}} 
\newcommand{\E}[1]{\*10^{#1}} 
\AtBeginDocument{\renewcommand{\u}[1]{\,\mathrm{#1}}} 

\AtBeginDocument{\renewcommand{\d}{\mr d}} 
\newcommand{\e}{\mr e} 

\newcommand{\nn}{\nonumber} 

\graphicspath{{Figures/}}

\title{Primordial black hole isocurvature modes from non-Gaussianity}
\author[a,b]{Rapha\"el van Laak}
\author[b]{Sam Young}
\affiliation[a]{Institute of Physics, \' Ecole Polytechnique F\' ed\' erale de Lausanne (EPFL),\\ BSP 617, Rte de la Sorge, CH-1015 Lausanne, Switzerland}
\affiliation[b]{Instituut-Lorentz for Theoretical Physics, Leiden University,\\Niels Bohrweg 2, 2333 CA Leiden, The Netherlands}

\emailAdd{raphael.vanlaak@epfl.ch}
\emailAdd{young@lorentz.leidenuniv.nl}


\arxivnumber{2303.05248}

\abstract{
Primordial black holes (PBHs) are black holes that might have formed in high density regions in the early universe.
The presence of local-type non-Gaussianity can lead to large-scale fluctuations in the PBH formation rate.
If PBHs make up a non-negligible fraction of dark matter, these fluctuations can appear as isocurvature modes, and be used to constrain the amplitude of non-Gaussianity.
Assuming that the parameters of non-Gaussianity are constant over all scales, we build upon the results of previous work by extending the calculation to include peaks theory and making use of the compaction $C$ for the formation criteria, accounting for non-linearities between $C$ and the curvature perturbation $\zeta$.
For quadratic models of non-Gaussianity, our updated calculation gives constraints that are largely unaltered compared to those previously found, while for cubic models the constraints worsen significantly.
In case all of the DM is made up of PBHs, the parameters of non-Gaussianity are $-2.9\E{-4}<f<3.8\E{-4}$ and $-1.5\E{-3}<g<1.9\E{-3}$ for quadratic and cubic models respectively.
}

\begin{document}

\maketitle
\flushbottom

\section{Introduction}
Primordial Black Holes (PBHs) are black holes that could have formed in the early universe in the radiation dominated regime after inflation.
PBHs are capable of providing an explanation for numerous observed and unexplained cosmological phenomena \cite{Clesse:2017bsw}.
Most relevantly, PBHs could (fully or partially) make up dark matter (DM) (see e.g. ref.~\cite{DMoverview} for an overview), be the seeds for supermassive black holes in the centres of galaxies~\cite{SMBHs} and be the source of observed LIGO/Virgo events~\cite{LIGOVIRGO}.

There are many mechanisms that could explain how PBHs are formed, including through the collapse of; cosmic strings loops~\cite{strings1,strings2,strings3,strings4,strings5,strings6}, bubble collisions~\cite{bubble1,bubble2,bubble3,bubble4,bubble5}, scalar field~\cite{scalar}, domain walls~\cite{wall1,wall2,wall3}, single-field inflation~\cite{singl1,singl2,singl3,singl4,singl5,Kawai:2021edk} and multi-field inflation~\cite{mingfl1,mingfl2,Kawai:2022emp}.
The most relevant production mechanism for this work however is through the collapse of regions with a large density during radiation domination.

Regions where the density exceeds a critical value as the density perturbation enters the horizon collapse and form a PBH.
This threshold value was first studied by Carr and Hawking~\cite{densitycollapse} who considered that gravity dominates over pressure when the length scale of the density fluctuation exceeds the Jeans length.
The jeans length $R_\mr J$ and the horizon length $R_\mr H$ are related by $R_\mr J/R_\mr H=\sqrt w$, with $w$ the equation of state parameter relating the pressure $p$ and density $\rho$ by $p=w\rho c^2$.
It takes the value $w=1/3$ in the radiation dominated regime.
It was found in ref.~\cite{Carr3} that for the region to collapse, the density contrast $\delta\equiv\delta\rho/\rho$, with $\delta\rho\equiv\rho-\rho_\mr b$ the difference between the density and the background density $\rho_\mr b$, should exceed a critical value of $\delta_\mr c\approx w$ at horizon crossing.

Many papers have since considered PBH formation (e.g.~\cite{deltac1,deltac2,deltac3,deltac4,deltac5,deltac6,YoungMuscoByrnes,Niemeyer1}), making use of both numerical and analytical methods, finding that the critical value is closer to $\delta_\mr c\approx0.5$ (where the exact value depends on the specific profile shape of the perturbation), and finding also that the PBH mass follows a scaling law given by
\begin{align}
M_\mr{PBH}=\kappa M_\mr H(\delta-\delta_\mr c)^\gamma,
\label{eq: mass scaling density}
\end{align}
with $\kappa\approx4$ and $\gamma \approx 0.36$ (where, again, the exact values can vary with the profile shape). The PBH mass therefore depends on the scale of the perturbation (described by the horizon mass, $M_\mr H$) and amplitude $\delta$ of the perturbation which forms a PBH.

The PBH mass is therefore of the same order as the horizon mass (see e.g. ref.~\cite{Carr})
\begin{align}
M_\mr H=\f{c^3t}{G}\sim10^{15}\bp{\f{t}{10^{-23}\u s}}\u g,
\label{eq: MPBH=MH}
\end{align}
where we have substituted the Hubble horizon mass in the radiation dominated regime in terms of cosmic time $t$, the speed of light $c$ and Newton's gravitational constant $G$.
Thus, depending on the time at which the PBH forms, the mass can be arbitrarily small or large.

The abundance of PBHs is described by the parameter
\begin{align}
\beta\equiv\left.\f{\rho_\mr{PBH}}{\rho}\right|_{t_\mr f},
\label{eq: beta def}
\end{align}
with $\rho_\mr{PBH}$ the PBH density and $\rho$ the density of the universe.
Both are evaluated at the time of PBH formation $t_\mr f$.
Cosmological observations place constraints on the abundance of PBHs (see e.g. refs.~\cite{Carr,DMoverview} and references therein).
These constraints usually give an upper bound on the fraction of DM that is made up of PBHs of a single mass $M$
\begin{align}
f_\mr{PBH}(M)\equiv\f{\Omega_\mr{PBH}(M)}{\Omega_\mr{DM}},
\label{eq: fPBH def}
\end{align}
where $\Omega_\mr{PBH}\equiv\rho_\mr{PBH}/\rho$ and $\Omega_\mr{DM}\equiv\rho_\mr{DM}/\rho$ the density parameters of PBHs and DM, respectively.

The abundance of PBHs can be strongly affected by primordial non-Gaussianity in the density distribution~\cite{nonGaussPBHs1,nonGaussPBHs2,nonGaussPBHs3,Ferrante:2022mui,Matsubara:2022nbr}.
It is therefore important to take non-Gaussianity into account when calculating PBH abundances, but we can also use this tight dependency to find constraints on the parameters that describe non-Gaussianity.
Previous work~\cite{C1} has done this by studying the effects of a peak-background split on modal coupling and constraints on the iso-curvature perturbations of the \emph{Planck Collaboration}~\cite{Planck}, which gave constraints on the relation between the fraction $f_\mr{PBH}$ and the parameters of non-Gaussianity.
Integral to this calculation was the use of Press-Schechter theory to calculate the abundance of PBHs that form in regions where the density perturbation exceeds the critical value.

In this paper, we will take into account recent developments in the field to calculate constraints more accurately (as described in e.g. ref.~\cite{C2}).
These developments suggest the use of peaks theory and the compaction (instead of Press-Schechter theory and the density contrast), as well as accounting for the mass-scaling relationship of PBHs.

This paper is structured as follows:
in section \ref{sec: Non-Gaussianity and primordial black holes} we give a brief overview of non-Gaussianity and the calculation of PBH abundance.
In section \ref{sec: Compaction and peaks theory} we describe how the abundance of PBHs can be evaluated.
In section \ref{sec: Constraints on the primordial black hole abundance} we derive up to date constraints on the abundance of PBHs, before summarising our findings in section \ref{sec: Conclusion}.

\section{Non-Gaussianity and primordial black holes}
\label{sec: Non-Gaussianity and primordial black holes}
\subsection{Non-Gaussianity}
Primordial fluctuations created during inflation are predicted and observed to follow a distribution that is very close to Gaussian on the CMB scales~\cite{WMAP}.
However, many models of inflation predict the emergence of some levels of non-Gaussianity (see e.g. ref.~\cite{nonGauss2}).
The detection of Non-Gaussianity therefore offers important insights into the workings of inflation.
The presence of even small amounts of primordial non-Gaussianity can also have a large impact on the amount of PBHs that are formed~\cite{nonGaussPBHs1,nonGaussPBHs2,nonGaussPBHs3,Ferrante:2022mui,Matsubara:2022nbr}.

The curvature perturbation $\zeta$ appears as a perturbative quantity in the FLRW metric
\begin{align}
\d s^2=-c^2\d t^2+\e^{2\zeta}a^2\bp{\d x^2+\d y^2+\d z^2},
\end{align}
with $a$ the scale factor.
Local-type non-Gaussianity can be modeled by expanding the curvature perturbation as a polynomial of a Gaussian distributed curvature perturbation $\zeta_\mr G$
\begin{align}
\zeta=\zeta_\mr G+f\bp{\zeta_\mr G^2-\sigma^2}+g\zeta_\mr G^3+\mc{O}\bp{\zeta_\mr{G}^4},
\label{eq: zeta transformation}
\end{align}
where $f=3f_\mr{NL,local}/5$ and $g=9g_\mr{NL,local}/25$ describe the amplitude of non-Gaussianity.
In eq. \eqref{eq: zeta transformation} the variance $\sigma^2=\ba{\zeta_\mr G^2}$ is subtracted to ensure that the average value $\ba\zeta$ vanishes.
Because the abundance of PBHs is sensitive to non-Gaussianity, it is important to take non-Gaussianity into account when calculating PBH abundances.
However, as we will see later, this can also result in strong constraints on the non-Gaussianity parameters if PBHs make up a non-negligible fraction of dark matter.

\subsection{Primordial black hole formation and the effect of non-Gaussianity}
PBHs do not form at a single time and the PBH density parameter can be expressed as~\cite{YoungMuscoByrnes,OmegaPBH}
\begin{align}
\Omega_\mr{PBH}=\int\limits_{M_\mr{min}}^{M_\mr{max}} \d\ln(M_\mr H)\,\bp{\f{M_\mr{eq}}{M_\mr H}}^{1/2}\beta(M_\mr H),
\end{align}
with $M_\mr{min}$ and $M_\mr{max}$ the smallest and largest horizon masses at which PBHs form and $M_\mr{eq}$ the horizon mass at matter-radiation equality. Note that this equation assumes pure radiation domination right up until matter-radiation equality.
The term $\bp{M_\mr{eq}/M_\mr H}^{1/2}\propto1/a_\mr H$ accounts for the redshift of the PBH density parameter during radiation domination, where $a_\mr H$ is the scale factor at the time of PBH formation.
With the horizon mass serving as a parameter of time, this integral integrates the PBH abundance over the whole period in which PBHs form.

Assuming that PBHs form over a short time-interval, parameterised by a single horizon mass, a good estimate of the PBH density parameter can be found as
\begin{align}
\Omega_\mr{PBH}\sim\bp{\f{M_\mr{eq}}{M_\mr H}}^{1/2}\beta(M_\mr H).
\label{eq: OmegaPBH approx}
\end{align}
This assumption can be valid in the case that the power spectrum peaks sharply at this scale (which will be assumed throughout this paper) --- meaning that PBH formation at other scales is negligible. This also coincides with the condition that the local-type expansion, equation \ref{eq: zeta transformation}, gives a valid description of the statistics of the compaction (see ref. \cite{Ferrante:2022mui} for more discussion).

We use $M_\mr{eq}=2.8\E{17}M_\odot$~\cite{YoungMuscoByrnes,Meq} and choose $M_\mr{H}=M_\odot$ throughout this paper, and note that the final constraints on the non-Gaussianity parameters depend only very weakly on this choice (see also the discussion in ref.~\cite{C1}), justifying also the approximation done in eq. \eqref{eq: OmegaPBH approx}.
Using the definition of the fraction $f_\mr{PBH}$, eq. \eqref{eq: fPBH def}, and $\Omega_\mr{CDM}=0.26$~\cite{OmegaCDM}, we then find
\begin{align}
\beta\sim\bp{\f{M_\mr \odot}{M_\mr{eq}}}^{1/2}\Omega_\mr{CDM}f_\mr{PBH}\sim10^{-10}f_\mr{PBH}.
\label{eq: beta approx fPBH}
\end{align}

In addition to the already mentioned effect on the total abundance of PBHs~\cite{nonGaussPBHs1,nonGaussPBHs2,nonGaussPBHs3,Ferrante:2022mui,Matsubara:2022nbr}, non-Gaussianity can also lead to the production of isocurvature modes in the early universe. The modal coupling which can arise as a result of the non-Gaussianity means that the amplitude of small-scale perturbations can be coupled to a long wavelength perturbation. This means that PBH formation can be enhanced (reduced) in regions where the small-scale perturbations are larger (smaller). These fluctuations in the PBH formation rate therefore appear as dark matter isocurvature perturbations. Refs.~\cite{C1,bias} calculated the amplitude of these isocurvature perturbations, and used costraints from the \emph{Planck Collaboration}~\cite{Planck} on isocurvature modes to find constraints on the non-Gaussianity parameters (as a function of the PBH abundance). The authors used a Press-Schechter approach, and applied the statistics of the curvature perturbation, derived from eq. \eqref{eq: zeta transformation},  (see e.g. ref.~\cite{nonGaussPBHs2}).

It has since been argued that using peaks theory is more suitable for calculating the PBH abundance~\cite{pk,pk1, pk2}.
As with Press-Schechter theory, peaks theory assumes that PBHs form in regions where perturbations exceed a critical value, but it also introduces the condition that PBHs form in regions where the perturbations are at a maximum.
Additionally, instead of using the curvature perturbation or density contrast to describe when a region forms a PBH, refs. ~\cite{Compaction1,Compaction2} argue that the compaction
\begin{align}
C(\mb x,r)\equiv2\f{M(\mb x,r,t)-M_\mr b(\mb x,r,t)}{R(\mb x,r,t)},
\label{eq: compaction}
\end{align}
is a more appropriate parameter use.
In the above, $M(\mb x,r,t)$ is the Misner-Sharp mass, with $M_\mr b(\mb x,r,t)$ its background value.
The Misner-Sharp mass gives the mass within a sphere of areal radius $R(\mb x,r,t)=a(t)\exp(\zeta(\mb x))r$ with spherical coordinate radius $r$, centred around position $\mb x$ and evaluated at time $t$. Refs.~\cite{C2,Ferrante:2022mui} have recently considered the effect of local-type non-Gaussianity on the statistics of the compaction, and we will apply their methods here.

As with the density contrast, PBHs form in regions where the compaction exceed a critical value.
Furthermore, like the density contrast, the compaction directly measures the overabundance of mass in a region and is therefore better suited than the curvature perturbation for determining when a region collapses.
The compaction is time-independent on super-horizon scales, and can also be expressed as the time-independent component of the top-hat smoothed density contrast $\delta_\mathrm{TH}$~\cite{Compaction2}
\begin{align}
\delta_{\rm{TH}}=\epsilon(t)^2C(\mb x,r),
\end{align}
where $\epsilon=r/r_\mathrm H$, the ratio between the perturbation scale $r$ and the Hubble scale $r_\mathrm{H}$. See ref. \cite{Compaction2} for more discussion on the compaction and its use as the formation criterion.

In light of these developments, ref.~\cite{C2} has evaluated the abundance of PBHs in the presence of local-type non-Gaussianity using peaks theory and the compaction, in addition to using the correct mass scaling in eq. \eqref{eq: mass scaling density}.
In this work, we will use these same methods to calculate the PBH abundance and use these methods to update the constraints on non-Gaussianity parameters previously found in ref.~\cite{C1} from the isocurvature modes.
For clarity, we will refer to the results of ref.~\cite{C1} as the results from \emph{Young and Byrnes}.

\section{Compaction and peaks theory}
\label{sec: Compaction and peaks theory}
In this section we will discuss how using the compaction, eq. \eqref{eq: compaction}, can be used to determine the abundance of PBHs.
We will follow the discussion in ref.~\cite{C2} in this section.
\subsection{Compaction}
\label{sec: pk compaction}

The compaction has a form similar to the density contrast.
Indeed, the compaction can be written in terms of the density contrast as
\begin{align}
C(\mb x,r)=\f{2}{R(r,t)}\int \d^3\mb x\,\rho_\mr b\delta(\mb x,t),
\label{eq: C integral}
\end{align}
where the integral of the density contrast over volume evaluates precisely the mass difference in eq. \eqref{eq: compaction}.
On super-horizon scales, the density contrast $\delta$ is related to the curvature perturbation in real space by the non-linear relation
\begin{align}
\delta(\mb x,t)=-\f{2(1+w)}{5+3w}\bp{\f{1}{aH}}^2e^{-2\zeta(\mb x)}\bp{\nabla^2\zeta(\mb x)+\f12(\vg\nabla\zeta(\mb x))^2}.
\label{eq: density contrast ur}
\end{align}
We shall take the equation of state parameter to be $w=1/3$ for the radiation dominated regime, and will take the high peak limit throughout.
The high-peak limit assumes that PBHs form in the high positive tail of the density distribution (or correspondingly, that peaks in the density fluctuation must have a high amplitude for PBH formation).
In the high peak limit, perturbations can be approximated as spherically symmetric \cite{pk}, and we find
\begin{align}
\delta(r,t)=-\f49\bp{\f{1}{aH}}^2e^{-2\zeta(r)}\bp{\zeta''(r)+\f2r\zeta'(r)+\f12\zeta'(r)^2},
\end{align}
where $\zeta'\equiv \d\zeta/\d r$.
Eq. \eqref{eq: C integral} then becomes
\begin{align}
C=\f{8\pi}{R(r,t)}\rho_\mr b\int_0^R\d\widetilde R\,\widetilde R(r,t)^2\delta(r,t)=-\f43r\zeta'(r)\bp{1+\f12r\zeta'(r)},
\end{align}
having used $R=a\exp(\zeta)r$.
Taking the linear component $C_1=-4r\zeta'(r)/3$, this gives the relation
\begin{align}
C=C_1-\f38C_1^2.
\label{eq: C}
\end{align}

This quadratic equation is sketched in figure \ref {fig: type I IIrange} and has a maximum at $C_1=4/3\equiv C_\mr{1,to}$, which we define as a turnover point, denoted by the subscript `to'.
The corresponding value of $C$ is $C_\mr{to}\equiv C(C_\mr{1,to})=2/3$.
Perturbations with $C_1<C_\mr{1,to}$ are referred to as type I perturbations, while perturbations with $C_1>C_\mr{1,to}$ are referred to as type II perturbations.
Type II perturbations are exponentially suppressed compared to type I perturbations as we will later see in eq. \eqref{eq: beta full} and section \ref{sec: peaks theory}, and the formation mechanism of PBHs from type II perturbations is not well understood~\cite{C2,TypeII}.
We will therefore limit our discussion to PBHs formed from type I perturbations.

\subsection{The effect of non-Gaussianity on the compaction}
\label{sec: pk non-Gauss}
The effect of non-Gaussianity on the compaction can be described by consulting eq. \eqref{eq: zeta transformation} to find that in the presence of non-Gaussianity, the linear term of the compaction becomes
\begin{align}
C_1=-\f43r\zeta'(r)=-\f43r\zeta_\mr G'(r)\bp{1+2f\zeta_\mr G(r)+3g\zeta_\mr G(r)^2}.
\label{eq: C1 ur}
\end{align}
The $-4r\zeta_\mr G'(r)/3$ term arises from the smoothing of the density contrast over a top-hat smoothing function, whilst the $\zeta_\mr G(r)$ term arises from the surface term, corresponding to a surface smoothing function. The top-hat and surface smoothing functions are defined, respectively, as
\begin{align}
W(\mb x,r)&\equiv\f{3}{4\pi r^3}\theta_\mr H(r-x),\\
W_\mr s(\mb x,r)&\equiv\f{1}{4\pi r^2}\delta_\mr D(x-r),
\end{align}
with $\theta_\mr H(x)$ the Heaviside step function and $\delta_\mr D(x)$ the Dirac delta function.
Their respective Fourier transforms are
\begin{align}
\widetilde W(k,r)&=3\f{\sin(kr)-kr\cos(kr)}{(kr)^3},\\
\widetilde W_\mr s(k,r)&=\f{\sin(kr)}{kr}.
\end{align}
We can use the smoothing functions to construct the following quantities which appear in the compaction, in eq. \eqref{eq: C1 ur}:
\begin{align}
C_\mr G(\mb x)=-\frac{4}{9}r^2\int \d^3\mb y\,\nabla^2\zeta_\mr G(\mb y)W(\mb x-\mb y,r)=-\f43r\zeta_\mr G'(r),
\end{align}
\begin{align}
\zeta_r(\mb x)=\int \d^3\mb y\,\zeta_\mr G(\mb y)W_\mr s(\mb x-\mb y,r)=\zeta_\mr G(r).
\end{align}
In each case, we have made the assumption of spherical symmetry for the second equality.

We also define the following correlation functions;
\begin{align}
\sigma_n^2&\equiv\f{16}{81}\int_0^k\f{\d k}{k}\,(kr)^4\widetilde W^2(k,r)k^{2n}P_{\zeta_\mr G},\label{eq: sigma n}\\
\sigma_r^2&\equiv\int_0^k\f{\d k}{k}\,\widetilde W_\mr s^2(k,r)P_{\zeta_\mr G},\\
\sigma_{0r}&\equiv\f49\int_0^k\f{\d k}{k}\widetilde W(k,r)\widetilde W_\mr s(k,r)P_{\zeta_\mr G},\label{eq: sigma 0r}
\end{align}
with $P_{\zeta_\mr G}$ the power spectrum of $\zeta_\mr G$.

To describe the PDF of $C_\mr G$ and $\zeta_r$, we define the following quantities, which are normalised to their variance,
\begin{align}
\nu&\equiv\f{C_\mr G}{\sigma_0},\\
\nu_r&\equiv\f{\zeta_r}{\sigma_r}.
\end{align}
The PDF is then given by the two-variate Gaussian distribution
\begin{align}
\mc N(\mb Y)\d\mb Y=\f{1}{\sqrt{2\pi\mr{det}(\Sigma)}}\exp\bp{-\f12\mb Y^\mr T\Sigma^{-1}\mb Y}\d\mb Y,
\end{align}
with $\mb Y=(\nu,\nu_r)$ and $\Sigma$ the covariance matrix.
To diagonalise this PDF, we introduce the variable
\begin{align}
z_r&\equiv\f{\nu_r-\gamma_{0r}\nu}{\sqrt{1-\gamma_{0r}^2}},
\end{align}
where
\begin{align}
\gamma_{0r}\equiv\f{\sigma_{0r}^2}{\sigma_0\sigma_r}
\label{eq: gamma0r}
\end{align}
is the correlation function of $\nu$ and $\nu_r$.
Like $\nu$ and $\nu_r$, the parameter $z_r$ follows a Gaussian distribution
\begin{align}
\mc N(z_r)\d z_r=\f{1}{\sqrt{2\pi(1-\gamma_{0r}^2)}}\exp\bp{-\f{(\nu_r-\gamma_{0r}\nu)^2}{2(1-\gamma_{0r}^2)}}\d \nu_r.
\label{eq: normal zr}
\end{align}
This gives the diagonalised PDF
\begin{align}
\mc N(\mb Y)\d\mb Y=\mc N(\nu)\mc N(z_r)\d\nu\d z_r=\f{1}{\sqrt{2\pi}}\exp\bp{-\f12\nu^2}\f{1}{\sqrt{2\pi}}\exp\bp{-\f12z_r^2}\d\nu\d z_r.
\end{align}

\begin{figure}[t]
\centering
\subfloat{\includegraphics[scale=0.6,valign=t]{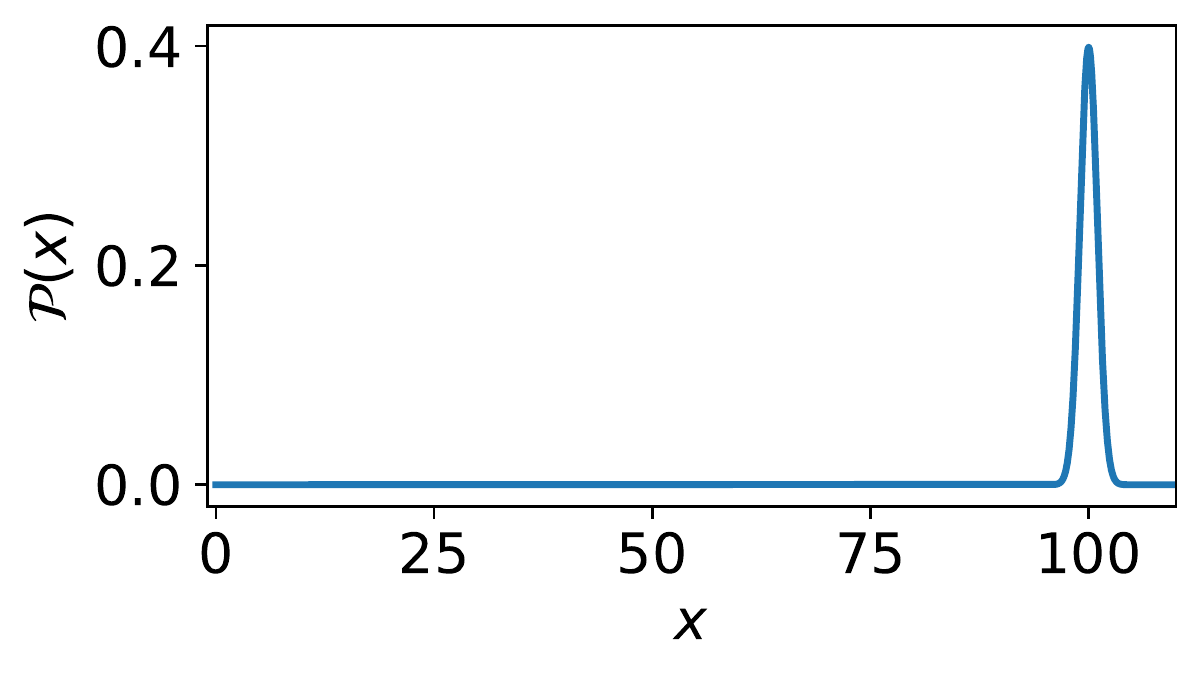}}%
\caption{The PDF of a Gaussian distribution where the average value is much larger than the standard deviation. Such distributions approach Dirac delta distributions.}
\label{fig: highpk}
\end{figure}

The high peak limit implies $\gamma_{0r}\nu\gg\sqrt{1-\gamma_{0r}^2}$.
This implies that the Gaussian distribution in eq. \eqref{eq: normal zr} has a mean that is much greater than its variance.
Such distributions can be estimated as Dirac delta functions (see figure \ref{fig: highpk}), and we therefore approximate
\begin{align}
\mc N(z_r)\d z_r\approx\delta_\mr D(\nu_r-\gamma_{0r}\nu)\d\nu_r.
\end{align}
Integrating the PDF over $\nu_r$ therefore results in the substitution $\nu_r=\gamma_{0r}\nu$, or equivalently
\begin{align}
\zeta_\mr G(r)=\gamma_{0r}\f{\sigma_r}{\sigma_0}C_\mr G.\label{eq: zetaG CG}
\end{align}
For brevity we define $\gamma\equiv\gamma_{0r}\sigma_r/\sigma_0$.
From eq. \eqref{eq: C1 ur} we then find
\begin{align}
C_1=C_\mr G+\widetilde{f}C_\mr G^2+\widetilde{g}C_\mr G^3,
\label{eq: C1}
\end{align}
where
\begin{align}
\widetilde{f}&\equiv2\gamma f,\\
\widetilde{g}&\equiv3\gamma^2g.
\end{align}

In this paper, to ensure that the local-type expansion considered is valid, we will consider only narrowly peaked power spectra, which can be well approximated by a Dirac delta function
\begin{align}
P_{\zeta_\mr G}(k)=\mc Ak\delta_\mr D(k-k_\mr p),
\label{eq: Dirac delta power spectrum}
\end{align}
with $\mc A$ the amplitude.
Although such a power law is unphysical, it justifies the use of the high peak limit and thus also spherical symmetry and is an accurate approximation of the lognormal power law often used in the literature~\cite{Compaction2,C2,Gow:2020bzo}.
Using this expression of the power law and eqs. \eqref{eq: sigma n} and \eqref{eq: sigma 0r} we can calculate
\begin{align}
\sigma_0^2&=\f{16}{81}(k_\mr pr)^4\widetilde W(k_p,r)^2\mc A\approx2.01\mc A,\\
\sigma_{0r}^2&=\f49(k_\mr pr)^2\widetilde W(k_p,r)\widetilde W_s(k_p,r)\mc A\approx0.200\mc A,
\end{align}
having used $r=2.74/k_p$, which relates the peak of the power spectrum to the amplitude scale (and maximises $\sigma_0$) \cite{Compaction2}.
Using eq. \eqref{eq: gamma0r}, we then also find
\begin{align}
\gamma=\gamma_{0r}\f{\sigma_r}{\sigma_0}\approx0.0995,
\end{align}
which is actually independent of $\sigma_r$.

\subsection{Primordial black hole abundance in peaks theory}
\label{sec: peaks theory}
\emph{Young and Byrnes} used Press-Schechter theory to calculate the abundance of PBHs, which assumes that PBHs form at points where the perturbation exceeds a critical value.
Peaks theory expends on this condition by adding the requirement that PBHs form at the point where the perturbation is at a local maximum.

The number density of peaks can be found to be~\cite{YoungMuscoByrnes}
\begin{align}
n_\mr{pk}(C_\mr G)=\f{1}{3^{3/2}(2\pi)^2}\bp{\f{\sigma_1}{\sigma_0}}^3\bp{\f{C_\mr G}{\sigma_0}}^3\exp\bp{\f{-C_\mr G^2}{2\sigma_0^2}}.
\label{eq: npk}
\end{align}

PBHs form in regions where the compaction exceeds a critical value $C_\mr{c}$.
The abundance of PBHs can be found as
\begin{align}
\beta=(2\pi)^{3/2}r^3\int \d C_\mr G\,\f{M_\mr{PBH}(C_\mr G)}{M_\mr H}n_\mr{pk}(C_\mr G).
\end{align}
We will use the mass scaling in eq. \eqref{eq: mass scaling density}, which in terms of the compaction reads
\begin{align}
M_\mr{PBH}(C)=KM_\mr H(C-C_\mr{c})^\gamma,
\end{align}
with $K=4$, $C_\mr{c}=0.5$ and $\gamma=0.36$~\cite{PBHparameters}.
Cf. eq. \eqref{eq: mass scaling density}, $C_\mr{c}$ is the critical value of PBH formation, i.e. PBHs form in regions where $C>C_\mr c$.

Combining these terms together with $\sigma_1/\sigma_0=k_p$ for the Dirac delta power spectrum \eqref{eq: Dirac delta power spectrum} in eq. \eqref{eq: sigma n} gives\footnote{We leave all expressions explicit in $C_\mr c$, but will always use $C_\mr c=0.5$.}
\begin{align}
\beta=\f{4\*2.74^3}{3^{3/2}\sqrt{2\pi}}\int \d C_\mr G\,(C-C_\mr c)^{0.36}\bp{\f{C_\mr G}{\sigma_0}}^3\exp\bp{-\f{C_\mr G^2}{2\sigma_0^2}}.
\label{eq: beta full}
\end{align}
Because $C_1$ takes larger values for type II perturbations than its does for type I perturbations, $C_\mr G$ will also take larger values for type II perturbations by eq. \eqref{eq: C1} for $\bv f,\bv g\lesssim0.1$.
The exponent in this integral then allows us to justify our earlier statement that type II perturbations are exponentially suppressed and therefore less relevant than type I perturbations.

\begin{figure}[t]
\centering
\subfloat{\includegraphics[scale=0.6,valign=t]{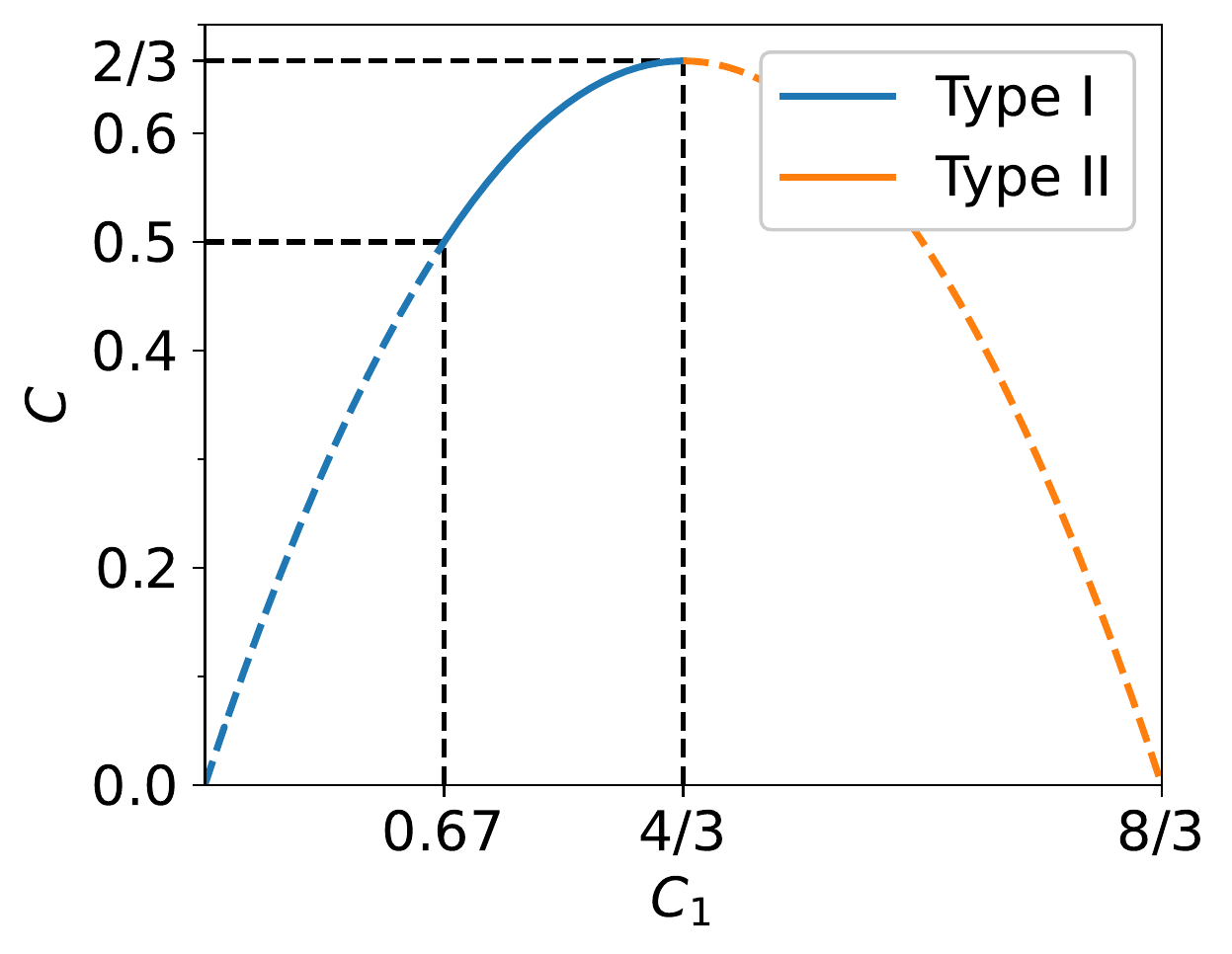}}%
\caption{Perturbations with $C_1<4/3$ are referred to as type I perturbations, while perturbations with $C_1>4/3$ are referred to as type II perturbations.
PBHs form when $C>C_\mr c$ and we only consider type I perturbations.
This means that we only consider the range $0.67=C_\mr{1,c}<C_1< C_\mr{1,to}=4/3$ in our calculations.}
\label{fig: type I IIrange}
\end{figure}

Using eq. \eqref{eq: C} and \eqref{eq: C1} for either a quadratic or cubic model of non-Gaussianity we can solve the above equation if the range of $C_\mr G$ is known.
PBHs form when the compaction exceeds the critical value, $C>C_\mr{c}$.
To find the value of $C_1$ corresponding to $C=C_\mr{c}$, we can invert eq. \eqref{eq: C} to find the solution
\begin{align}
C_1=\f43\bp{1-\sqrt{\f{2-3C}{2}}},
\end{align}
taking only solutions where $C_1<C_\mr{1,to}$, corresponding to type I perturbations.
For $C=C_\mr c=0.5$, this gives $C_1=0.67\equiv C_\mr{1,c}$.\footnote{Because the value $C_\mr c=0.5$ is not an exact solutions but a numerical result, we do not use the exact solution $2/3$ for $C_\mr{1,c}$.}
This means that PBHs form when
\begin{align}
0.67=C_\mr{1,c}<C_1< C_\mr{1,to}=\f43.
\end{align}
This corresponds to the range for $C_1$ where the compaction exceeds the critical value, but does not exceed the turnover point where perturbations become type II perturbations (see figure \ref{fig: type I IIrange}).

Because eq. \eqref{eq: beta full} contains an integral over $C_\mr G$ we need to invert eq. \eqref{eq: C1} to find the range of $C_\mr G$ that corresponds with the above range of $C_1$.
We will do this for the quadratic and cubic models separately.

\subsubsection{Quadratic expansion}
\label{sec: int range quadratic}
\begin{figure}[t]
\centering

\subfloat{\includegraphics[scale=0.6,valign=t]{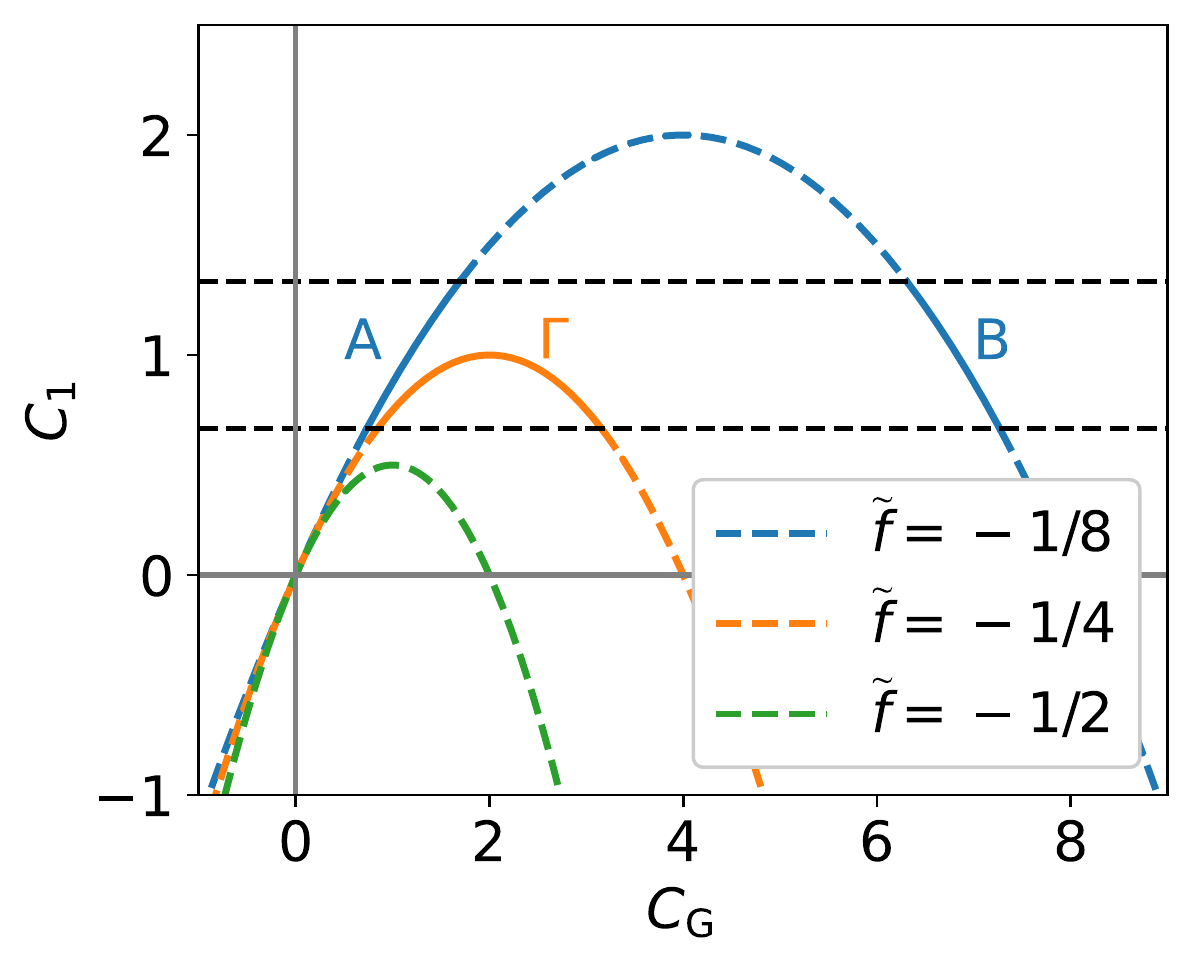}}%

\caption{Sketch of eq. \eqref{eq: C1 quadratic} for different values of $\widetilde{f}$. Indicated by the horizontal dashed lines are the lower and upper bounds of $C_1$, between which PBHs form from type I perturbations. Included are plots for which $\widetilde{f}>-1/(4C_\mr{1,to})$ (blue), $-1/(4C_\mr{1,c})<\widetilde{f}\leq-1/(4C_\mr{1,to})$ (orange) and $\widetilde{f}\leq-1/(4C_\mr{1,c})$ (green). We have indicated each range in which PBHs form with Greek letters as described in the text. Indicated by the dashed lines are $C_\mr{1,c}$ and $C_\mr{1,to}$.}
\label{fig: bounds quadratic}
\end{figure}

For a quadratic expansion (i.e. $g=0$) eq. \eqref{eq: C1} becomes
\begin{align}
C_1=C_\mr G+\widetilde{f}C_\mr G^2,
\label{eq: C1 quadratic}
\end{align}
which we can then invert to find two solutions
\begin{align}
C_\mr{G,\pm}(C_1)=\f{-1\pm\sqrt{1+4\widetilde{f}C_1}}{2\widetilde{f}}.
\end{align}
Multiple solutions are only possible when $1+4\widetilde{f}C_1>0$.
Because PBHs form when $C_\mr{1,c}<C_1<C_\mr{1,to}$ we can then identify the corresponding ranges over which $C_\mr G$ needs to be integrated in eq. \eqref{eq: beta full}.
These ranges depend on the value of $\widetilde{f}$ and we will indicated each of these ranges by Greek capital letters and include those in figure \ref{fig: bounds quadratic}.
\begin{itemize}
\item For $\widetilde{f}>-1/(4C_\mr{1,to})$ the range of integration is between $C_\mr{G,+}(C_\mr{1,c})<C_\mr G<C_\mr{G,+}(C_\mr{1,to})$ (A) and between $C_\mr{G,-}(C_\mr{1,to})<C_\mr G<C_\mr{G,-}(C_\mr{1,c})$ (B).
\item For $-1/(4C_\mr{1,c})<\widetilde{f}\leq-1/(4C_\mr{1,to})$ the range of integration is between $C_\mr{G,+}(C_\mr{1,c})<C_\mr G<C_\mr{G,-}(C_\mr{1,c})$ ($\Gamma$).
\item For $\widetilde{f}\leq-1/(4C_\mr{1,c})$ PBHs do not form.
\end{itemize}

\subsubsection{Cubic expansion}
\label{sec: int range cubic}
\begin{figure}[t]
\centering
\subfloat{\includegraphics[scale=0.6,valign=t]{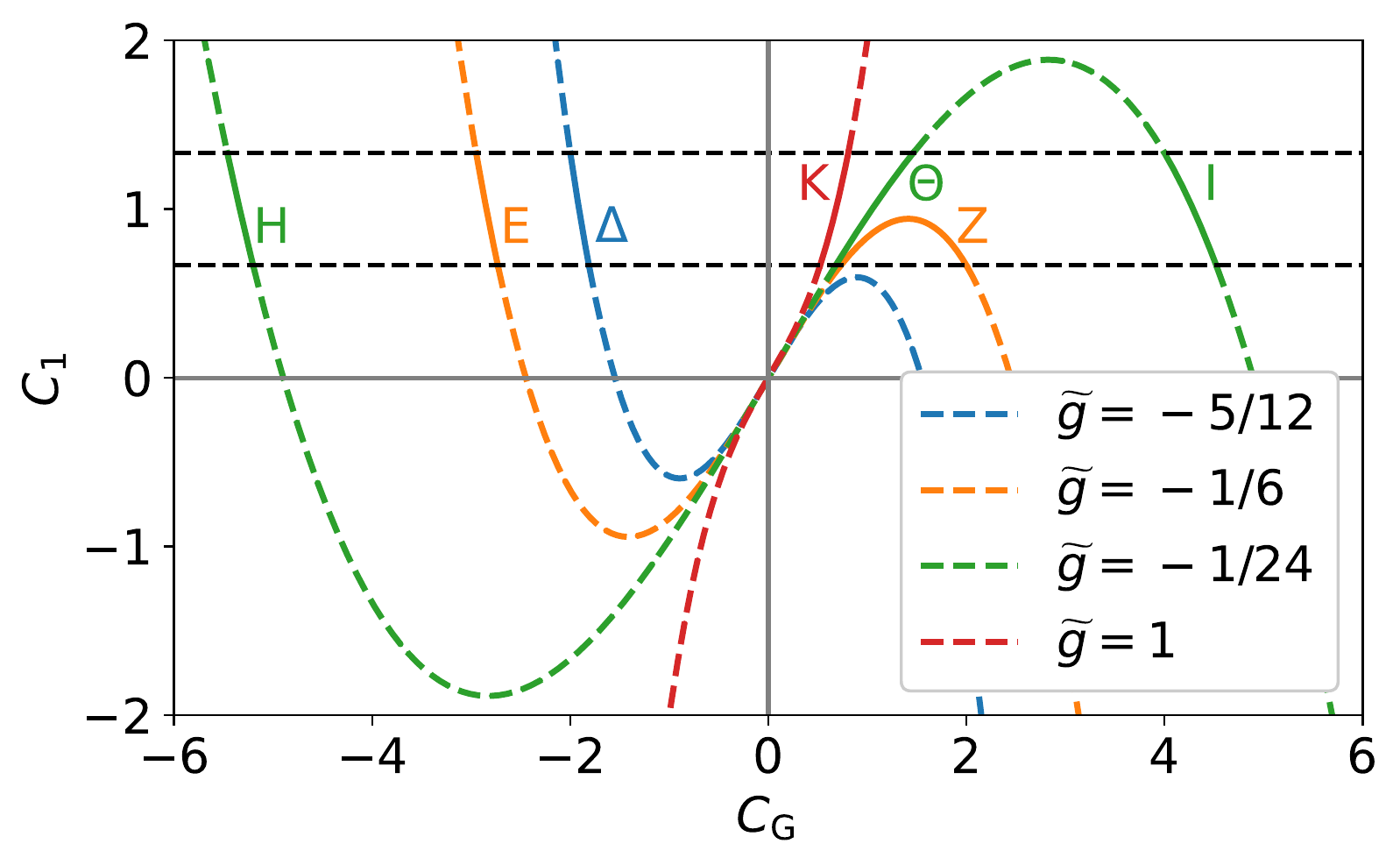}}%
\caption{Sketch of eq. \eqref{eq: C1 cubic} for different values of $\widetilde{g}$. Indicated by the horizontal dashed lines are the lower and upper bounds of $C_1$, between which PBHs form from type I perturbations. Included are plots for which $\widetilde{g}\leq-4/(27C_\mr{1,c}^2)$ (blue), $-4/(27C_\mr{1,c}^2)<\widetilde{g}\leq-4/(27C_\mr{1,to}^2)$ (orange), $-4/(27C_\mr{1,to}^2)<\widetilde{g}\leq0$ (green) and $\widetilde{g}>0$ (red). We have indicated each range in which PBHs form with Greek letters as described in the text. Indicated by the dashed lines are $C_\mr{1,c}$ and $C_\mr{1,to}$.}
\label{fig: bounds cubic}
\end{figure}

For a cubic expansion (i.e. $f=0$), eq. \eqref{eq: C1} becomes
\begin{align}
C_1=C_\mr G+\widetilde{g}C_\mr G^3,
\label{eq: C1 cubic}
\end{align}
which we can invert to find three solutions
\begin{align}
C_{\mr G,i}(C_1)=\f{Q_i}{2^{2/3}3^{1/3}\lambda(C_1)}-\f{Q_i^*\lambda(C_1)}{2^{1/3}3^{2/3}\widetilde{g}},
\end{align}
with
\begin{align}
Q_i&=(-2,1-i\sqrt3,1+i\sqrt3),\\
\lambda(C_1)&=\bp{9C_1\widetilde{g}^2+\sqrt3\sqrt{4\widetilde{g}^3+27C_1^2\widetilde{g}^4}}^{1/3}.
\end{align}
Similar to before, the range of integration depends on the value of $\widetilde{g}$.
The general behaviour again depends on the square root term.
However, because $Q_i$ could also be complex, we no longer necessarily require $4\widetilde{g}^3+27C_1^2\widetilde{g}^4>0$.
As before, we indicate all ranges of $C_\mr G$ with Greek letters and include them in figure \ref{fig: bounds cubic}.
\begin{itemize}
\item For $\widetilde{g}\leq-4/(27C_\mr{1,c}^2)$ the range of integration is $C_\mr{G,1}(C_\mr{1,to})<C_\mr G<C_\mr{G,1}(C_\mr{1,c})$ ($\Delta$).
\item For $-4/(27C_\mr{1,c}^2)<\widetilde{g}\leq-4/(27C_\mr{1,to}^2)$ the range of integration is $C_\mr{G,1}(C_\mr{1,to})<C_\mr G<C_\mr{G,1}(C_\mr{1,c})$ (E), as well as $C_\mr{G,2}(C_\mr{1,c})<C_\mr G<C_\mr{G,3}(C_\mr{1,c})$ (Z).
\item For $-4/(27C_\mr{1,to}^2)<\widetilde{g}\leq0$ the range of integration is $C_\mr{G,1}(C_\mr{1,to})<C_\mr G<C_\mr{G,1}(C_\mr{1,c})$ (H), as well as $C_\mr{G,2}(C_\mr{1,c})<C_\mr G<C_\mr{G,2}(C_\mr{1,to})$ ($\Theta$) and $C_\mr{G,3}(C_\mr{1,to})<C_\mr G<C_\mr{G,3}(C_\mr{1,c})$ (I).
\item For $\widetilde{g}>0$ the range of integration is $C_\mr{G,1}(C_\mr{1,c})<C_\mr G<C_\mr{G,1}(C_\mr{1,to})$ (K).
\end{itemize}

\section{Constraints on the primordial black hole abundance}
\label{sec: Constraints on the primordial black hole abundance}
\subsection{Peak-background split}
\label{eq: pk: background split}
\begin{figure}[t]
\centering
\includegraphics[scale=0.6,valign=t]{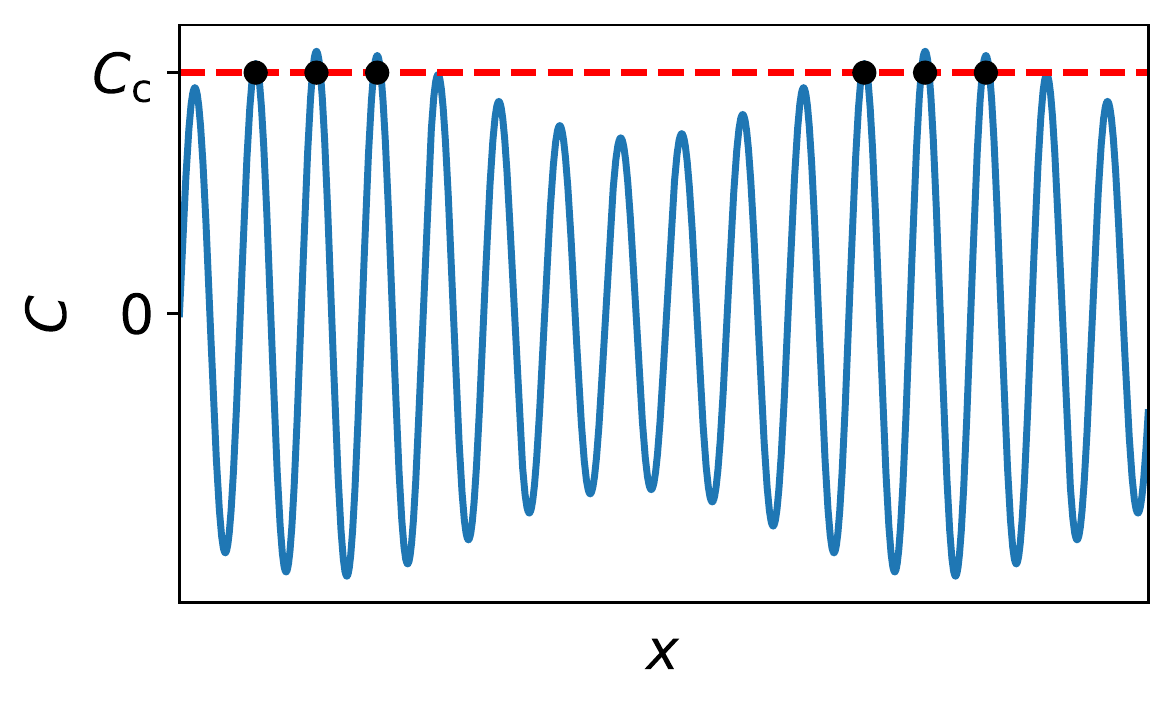}
\caption{The effect of modal coupling. Normally, the amplitude of a short wavelength mode would not exceed the critical value of PBH formation. However, when a long wavelength mode is present in the background, the amplitude of short wavelength modes could be boosted to above the critical value. In this figure, we have subtracted terms that depend exclusively on long wavelength modes, similar to eq. \eqref{eq: quadratic} and \eqref{eq: cubic}.}
\label{fig: modal coupling}
\end{figure}

Under a peak-background split, perturbations could be split into a small-scale ``peak'' component $\zeta_\mr s$ and a large-scale ``background'' component $\zeta_\mr l$
\begin{align}
\zeta_\mr G=\zeta_\mr s+\zeta_\mr l.
\label{eq: s/l expansion}
\end{align}
Up to first order in Fourier space, eq. \eqref{eq: density contrast ur} becomes
\begin{align}
\delta(\mb k,t)=\f{2(1+w)}{5+3w}\bp{\f{k}{aH}}^2\zeta(\mb k).
\end{align}
Because large-scale fluctuations are suppressed by a factor $k^2$, long wavelength perturbations do not contribute directly to PBH formation~\cite{zetac2}.
In the presence of local-type non-Gaussianity, they do however contribute to PBH formation indirectly.
The amplitude of small wavelengths will be boosted around peaks of long wavelengths, increasing the probability that the total perturbation exceeds the critical perturbation~\cite{C1} (see figure \ref{fig: modal coupling}).
Small-scale perturbations need to be much larger than those with CMB scale wavelengths in order for a significant number of PBHs to be formed.
Following the approach in \emph{Young and Byrnes}, we will assume $\zeta_\mr l\ll1$ and analyse the effect that this has on the abundance of PBHs under quadratic and cubic expansion of non-Gaussianity.

\subsection{Effects of peak-background split on the compaction}
We will now consider the effects of the peak background split, eq. \eqref{eq: s/l expansion}, on the compaction.

\subsubsection{Quadratic expansion}
\label{sec: peak split quadratic expansion}
Using eqs. \eqref{eq: C1 ur} and \eqref{eq: zeta transformation} for the linear component of the compaction for a quadratic model (i.e. $g=0$), we can add in the peak-background split using eq. \eqref{eq: s/l expansion}
\begin{align}
\nn C_1&= -\frac{4}{3} r \zeta'(r)\\
\nn    &= -\frac{4}{3} r \bs{ (\zeta_\mr s+\zeta_\mr l) + f \bp{(\zeta_\mr s+\zeta_\mr l)^2 -\ba{(\zeta_\mr s+\zeta_\mr l)^2}  }}'\\
\nn	&= -\frac{4}{3} r\bs{ \zeta_\mr s' + \cancel{\zeta_\mr l'} + 2 f \zeta_\mr s\zeta_\mr s' + \cancel{ 2f\zeta_\mr s\zeta_\mr l'} +  2f\zeta_\mr l\zeta_\mr s' + \cancel{ 2f\zeta_\mr l\zeta_\mr l'}}\\
 &=\bp{1+2f\zeta_\mr l}C_\mr G+\widetilde{f}C_\mr G^2,
 \label{eq: quadratic}
\end{align}
where in the final line we have neglected derivatives in $\zeta_\mr l$, substituted $C_\mr G = -4r \zeta_\mr s'/3$ and we have used eq. \eqref{eq: zetaG CG}, which implies that up to first order
\begin{align}
2f\zeta_\mr s=2f\gamma C_\mr G=\widetilde{f}C_G.
\end{align}
Eq. \eqref{eq: quadratic} can now be solved for $C_\mr G$ to give
\begin{align}
C_\mr{G,\pm}(C1)=\f{-(1+2f\zeta_\mr l)\pm\sqrt{(1+2f\zeta_\mr l)^2+4\widetilde{f}C_1}}{2\widetilde{f}}.
\end{align}
For $\zeta_\mr l\ll1$ the range of PBH formation for $C_\mr G$ will be very similar to those calculated in the previous section:
\begin{itemize}
\item For
\begin{align}
\f{4\widetilde{f}}{(1+2f\zeta_\mr l)^2}>-\f{1}{C_\mr{1,to}}
\end{align}
PBHs form in the ranges $C_\mr{G,+}(C_\mr{1,c})<C_\mr G<C_\mr{G,+}(C_\mr{1,to})$ and $C_\mr{G,-}(C_\mr{1,to})<C_\mr G<C_\mr{G,-}(C_\mr{1,c})$.
\item For
\begin{align}
-\f{1}{C_\mr{1,c}}<\f{4\widetilde{f}}{(1+2f\zeta_\mr l)^2}\leq-\f{1}{C_\mr{1,to}}
\end{align}
PBHs form in the range $C_\mr{G,+}(C_\mr{1,c})<C_\mr G<C_\mr{G,-}(C_\mr{1,c})$.
\item For
\begin{align}
\f{4\widetilde{f}}{(1+2f\zeta_\mr l)^2}\leq-\f{1}{C_\mr{1,c}}
\end{align}
PBHs do not form.
\end{itemize}

\subsubsection{Cubic expansion}
\label{sec: peak split cubic expansion}
When including only the linear and cubic terms in eqs. \eqref{eq: C1 ur} and \eqref{eq: zeta transformation} (i.e. $f=0$), the linear component of the compaction transforms as
\begin{align}
\nn C_1&=-\f43r\zeta'(r)\\
\nn&=-\f43r\bs{(\zeta_\mr s+\zeta_\mr l)+g\bp{\zeta_\mr s+\zeta_\mr l}^3}'\\
\nn&=-\f43r\bs{\zeta_\mr s'+\cancel{\zeta_\mr l'}+3g\zeta_\mr s^2\zeta_\mr s'+\cancel{3\zeta_\mr s^2\zeta_\mr l'}+6\zeta_\mr s\zeta_\mr l\zeta_\mr s'+\cancel{6\zeta_\mr s\zeta_\mr l\zeta_\mr l'}+3\zeta_\mr l^2\zeta_\mr s'+\cancel{3\zeta_\mr l^2\zeta_\mr l'}}'\\
&=\bp{1+3g\zeta_\mr l^2}C_\mr G+6g\gamma\zeta_\mr lC_\mr G^2+\widetilde{g}C_\mr G^3,
\label{eq: cubic}
\end{align}
having again neglected derivatives in $\zeta_\mr l$, substituted $C_\mr G=-4r\zeta_\mr s'/3$ and used eq. \eqref{eq: zetaG CG} to find
\begin{align}
3g\zeta_\mr s^2=3g\gamma^2C_\mr G^2=\widetilde{g}C_\mr G^2.
\end{align}
Similar to before, we can now solve eq. \eqref{eq: cubic} for $C_\mr G$ to find the range of integration for $C_\mr G$ corresponding to $C_\mr{1,c}<C_1<C_\mr{1,to}$.
Doing so gives three solutions $C_{\mr{G},i}(C_1)$, with $i\in\bc{1,2,3}$ similar to those discussed in section \ref{sec: int range cubic}.
However, even though the additional quadratic term in the above equation does not bring any new free parameters, it still brings significant complexity to the algebraic form of these solutions.
This makes it significantly harder to identify the range of integration.

However, since we can assume $\zeta_\mr l\ll1$, we can simplify the analysis and ignore two of the three solutions.
The reason for this depends on the sign of $g$:
\begin{itemize}
\item
When $g>0$ there is only one real solution $C_{\mr G,j}(C_1)$, for $C_\mr{1,c}<C_1<C_\mr{1,to}$ and $\zeta_\mr l\ll1$, with the other two solutions $C_{\mr G,k\neq j}(C_1)$ being complex.

\item
For $g<0$ there will always be 3 real solutions $C_{\mr G,i}(C_1)$, as long as $C_\mr{1,c}<C_1<C_\mr{1,to}$ and $g$ is sufficiently small in magnitude ($g\gtrsim-1$).
However, the smallest of these solutions $C_{\mr G,j}(C_1)$ is smaller than the other two $C_{\mr G,k\neq j}(C_1)$ by a factor $\sim$ few.
Because the integrand in eq. \eqref{eq: beta full} depends exponentially on $\bv{C_\mr G}$, contributions from larger values of $\bv{C_\mr G}$ will be exponentially suppressed.
We can therefore ignore these larger solutions.
\end{itemize}
In conclusion, irrespective of the sign of $g$, we can then integrate over the range $C_{\mr G,j}(C_\mr{1,c})<C_\mr G<C_{\mr G,j}(C_\mr{1,to})$.

\subsection{Bias factor and non-Gaussianity parameters}
\label{sec: pk theory Constraints on the non-Gaussianity parameters}
Having properly found the bounds of the integral in eq. \eqref{eq: beta full}, we can calculate the PBH abundance $\beta$.
Eqs. \eqref{eq: quadratic} and \eqref{eq: cubic} offer perturbations compared to eq. \eqref{eq: C1 ur} in terms of $\zeta_\mr l$.
We can use this to express the relative change in PBH abundance under the peak-background split as
\begin{align}
\delta_\beta=\f{\beta(\zeta_\mr l)-\beta(\zeta_\mr l=0)}{\beta(\zeta_\mr l=0)}.
\label{eq: delta beta}
\end{align}
In the above, $\beta$ is evaluated using eq. \eqref{eq: beta full}, which depends on the compaction $C$ and $C_\mr G$.
The compaction $C$ in turn depends on the linear component $C_1$ as in eq. \eqref{eq: C}, which depends on the Gaussian component $C_\mr G$ and the non-Gaussianity parameters $f$ and $g$ as in eqs. \eqref{eq: quadratic} and \eqref{eq: cubic} for quadratic and cubic expansions respectively.

For the term $\beta(\zeta_\mr l)$ in eq. \eqref{eq: delta beta}, the perturbed values of $C_\mr 1$ from eqs. \eqref{eq: quadratic} and \eqref{eq: cubic} are used, along with the bounds as found in sections \ref{sec: peak split quadratic expansion} and \ref{sec: peak split cubic expansion} to evaluate the integral.
For the background value term $\beta(\zeta_\mr l=0)$, one simply uses eq. \eqref{eq: C1} for the linear component $C_1$ and the integration ranges of $C_\mr G$ as described in sections \ref{sec: int range quadratic} and \ref{sec: int range cubic}.

\begin{figure}[t]
\centering
\subfloat{\includegraphics[scale=0.55,valign=t]{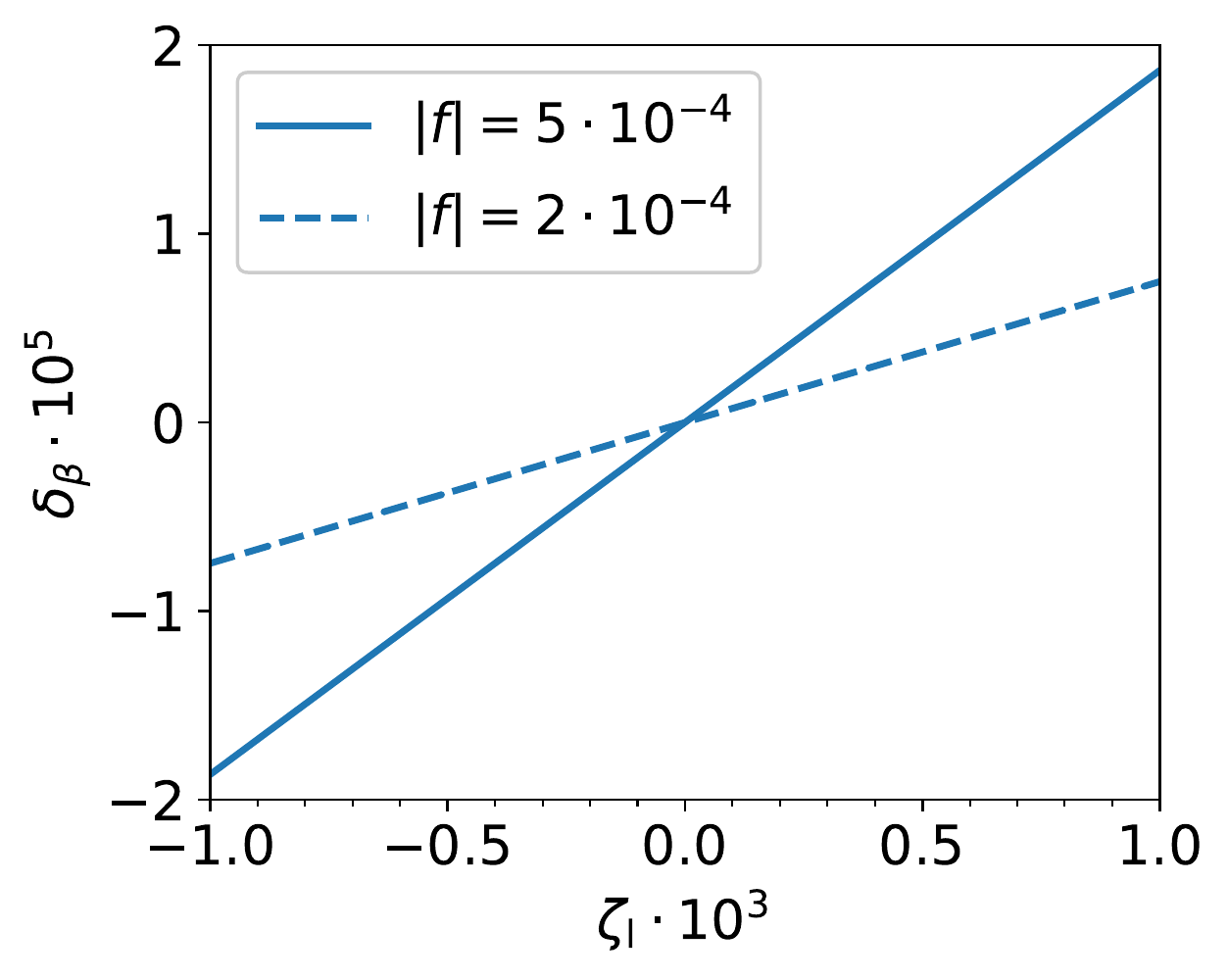}}%
\quad%
\subfloat{\includegraphics[scale=0.55,valign=t]{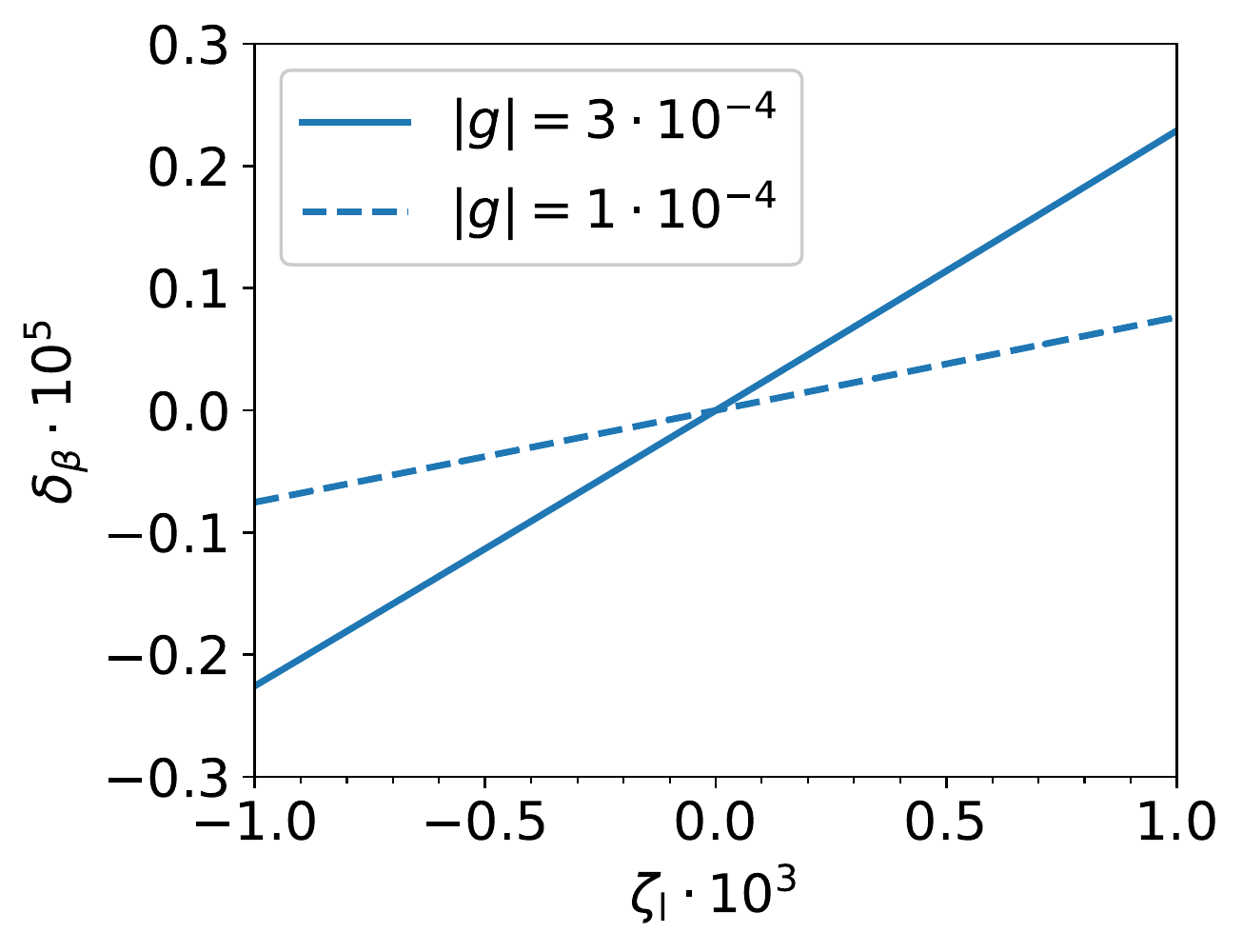}}
\caption{Graph of the change in $\delta_\beta$ as a function of small $\zeta_\mr l$ for different values of $f$ in the quadratic expansion (left panel) and $g$ for the cubic expansion (right panel).
We use $\sigma_0=0.15$ for reference here.}
\label{fig: delta beta_zeta l}
\end{figure}

Figure \ref{fig: delta beta_zeta l} shows $\delta_\beta$ as a function of $\zeta_\mr l$, and we see that, for the range of values considered, we can express $\delta_\beta$ as a linear function of $\zeta_l$
\begin{align}
\delta_\beta=b\zeta_\mr l,
\label{eq: bias factor}
\end{align}
with $b$ the bias factor~\cite{bias}.

The DM density parameter $\Omega_\mr{DM}$ up to first order in $\zeta$ is related to the background density parameter $\overline\Omega_\mr{DM}$ by
\begin{align}
\Omega_\mr{DM}=(1+f_\mr{PBH}b\zeta+3\zeta)\overline\Omega_\mr{DM}.
\end{align}
As the universe expands over time, matter perturbations evolve by a factor $(a\exp(\zeta))^{-3}$, which means that the $3\zeta$ term is the adiabatic mode to first order in $\zeta$.
The $f_\mr{PBH}b\zeta$ term is an isocurvature mode and forms a deviation from the adiabatic mode.
The isocurvature mode is either fully correlated or anti-correlated, depending on the sign of the non-Gaussianity parameters~\cite{C1}.
The \emph{Planck Collaboration}~\cite{Planck} has found constraints on isocurvature modes.
On CMB scales these are
\begin{align}
\beta_\mr{iso}=\begin{cases}1.3\E{-3},&\text{fully correlated},\\8\E{-4},&\text{fully anti-correlated},\end{cases}
\label{eq: Planck constraints}
\end{align}
with fully correlated modes corresponding to $b>0$ and fully anti-correlated modes corresponding to $b<0$.
We can express $\beta_\mr{iso}$ in the above as
\begin{align}
\beta_\mr{iso}=\f{P_\mr{iso}}{P_\mr{iso}+P_\zeta},
\end{align}
where $P_\mr{iso}$ is the isocurvature power spectrum and $P_\zeta$ the perturbation power spectrum.
These are related as~\cite{C1}
\begin{align}
P_\mr{iso}=b^2P_\zeta.
\end{align}
If a fraction $f_\mr{PBH}$ of DM is made up of PBHs, the constraints \eqref{eq: Planck constraints} can then be used to constrain $b$ as
\begin{align}
-0.028<bf_\mr{PBH}<0.036.
\label{eq: b constraints}
\end{align}

With $b$ being independent on $\zeta_\mr l$, there are three free parameters in this inequality; $f$ or $g$ for respectively quadratic and cubic expansion, $\sigma_0$, as it appears in eq. \eqref{eq: beta full}, and $f_\mr{PBH}$.
Together with eq. \eqref{eq: beta approx fPBH}, we can then find a relation between $f_\mr{PBH}$ and $f$ or $g$ by the system of equations
\begin{align}
\beta&=10^{-10}f_\mr{PBH},\label{eq: coupled equation1}\\
bf_\mr{PBH}&=\begin{cases}0.036,&b>0,\\-0.028,&b<0.\end{cases}
\label{eq: coupled equation2}
\end{align}
The sign of $b$ is the same as that of $f$ or $g$.
Thus, the condition of the sign of $b$ in eq. \eqref{eq: coupled equation2}, can also be understood as a condition on the sign of $f$ or $g$.

In eqs. \eqref{eq: coupled equation1} and \eqref{eq: coupled equation2}, both $\beta$ and $b$ depend on $\sigma_0$ and $f$ or $g$.
Thus, eq. \eqref{eq: coupled equation1} can be solved for $\sigma_0$ as a function of $f$ or $g$ and $f_\mr{PBH}$.
We can then insert this relation in eq. \eqref{eq: coupled equation2} for the dependency of $b$ on $\sigma_0$ and solve to find a relation between $f_\mr{PBH}$ and $f$ or $g$.
This relation gives constraints on $f$ or $g$ for a given value of $f_\mr{PBH}$.

\begin{figure}[t]
\centering
\subfloat{\includegraphics[scale=0.55,valign=t]{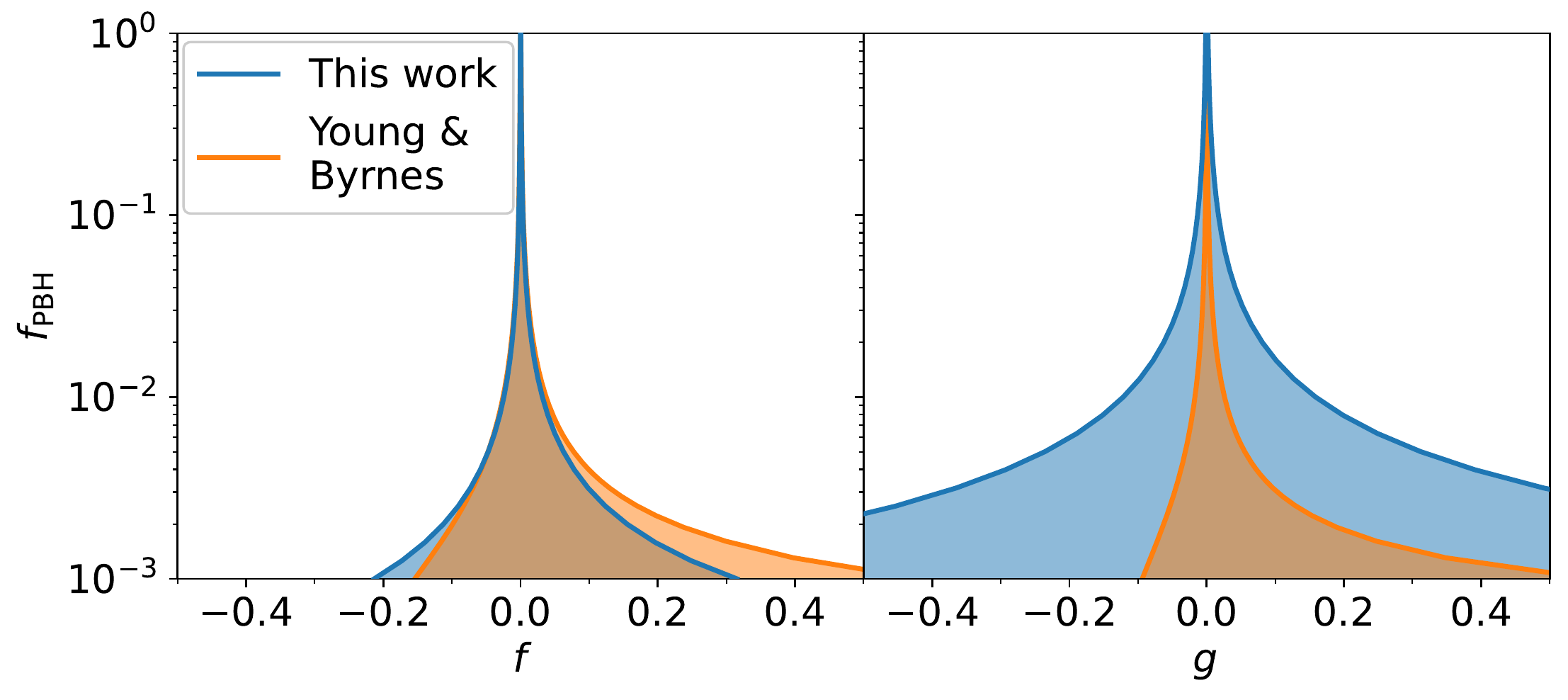}}%
\caption{Constraints on $f$ and $g$ for a given value of $f_\mr{PBH}$ as found in this work compared to those found by \emph{Young and Byrnes} for quadratic expansions (left panel) and cubic expansions (right panel).
Indicated are the allowed values for $f$ and $g$ for a given value of $f_\mr{pbh}$.}
\label{fig: rPBH}
\end{figure}

The resulting relation between $f_\mr{PBH}$ and $f$ or $g$ is shown in figure \ref{fig: rPBH}.
We compare these to the constraints that were found in \emph{Young and Byrnes}.
To plot the results from \emph{Young and Byrnes}, we followed the same approach as the one presented in their paper.

From the results in figure \ref{fig: rPBH} we find that if we set $f_\mr{PBH}=1$, which corresponds to all of the DM being made up of PBHs, we find the bounds
\begin{align}
-2.9\E{-4}<f<3.8\E{-4},\\
-1.5\E{-3}<g<1.9\E{-3},
\end{align}
for the quadratic and cubic models respectively.
The results for the quadratic model are similar to those found by \emph{Young and Byrnes}, but broader for the cubic model by order $\mc O(10)$.

We see that the constraints on $f$ found in this work remain mostly unaltered in both shape and magnitude compared to those found by \emph{Young and Byrnes}.
For a cubic expansion on the other hand the constraints look significantly different than before in two different ways.
Firstly, our results display significant broadening at all ranges of $g$, leading to much weaker constraints, and secondly, the constraints become a lot less symmetric. Both of these effects are explained below.

\paragraph{Broadening of constraints for cubic models.}
In this work, constraints based on isocurvature modes were found by expanding the linear component of the compaction $C_1$.
Meanwhile, \emph{Young and Byrnes} found the constraints by expanding the curvature perturbation $\zeta$ in $\zeta_\mr l$ instead.
We therefore compare the precise form of the transformation of $\zeta$ in the cubic model when performing a peak-background split in eq. \eqref{eq: zeta transformation}~\cite{C1}
\begin{align}
\zeta=\bp{1+3g\zeta_\mr l^2}\zeta_\mr s+3g\zeta_\mr l\zeta_\mr s^2+g\zeta_\mr s^3
\label{eq: zeta cubic reiteration}
\end{align}
with that of $C_1$, given by eq. \eqref{eq: cubic}
\begin{align}
C_1=\bp{1+3g\zeta_\mr l^2}C_\mr G+6g\gamma\zeta_\mr lC_\mr G^2+\widetilde{g}C_\mr G^3.
\label{eq: cubic reiteration}
\end{align}
We see that in eq. \eqref{eq: cubic reiteration} the linear term of $C_1$ is perturbed by $\mc O(\zeta_\mr l^2)$, the quadratic term by $\mc O(\zeta_\mr l)$, while the cubic term remains unperturbed.
Because $b$ is determined from perturbations in $\zeta_\mr l\ll1$, we can conclude that the quadratic term in eq. \eqref{eq: cubic reiteration} has the most dominant effect on $b$, and therefore also on $f_\mr{PBH}$ by eq. \eqref{eq: coupled equation2}.
The same could be said for the quadratic term in eq. \eqref{eq: zeta cubic reiteration}.
However, while the prefactor of the quadratic term in eq. \eqref{eq: zeta cubic reiteration} is $3g\zeta_\mr l$, that of the quadratic term in eq. \eqref{eq: cubic reiteration} is $6g\gamma\zeta_\mr l$ and is thus smaller by a factor $2\gamma=\mc O(0.1)$.

Because the constraints in our work relied on perturbations in $C_1$, while those found in \emph{Young and Byrnes} relied on perturbations in $\zeta$, this leads to a broadening of roughly magnitude $\mc O(10)$ for the constraints found in this work as we need $g$ to be larger by the same order $\mc O(10)$ to compensate for this smaller prefactor.
From figure \ref{fig: rPBH} we can indeed see that for lower values of $g$, the constraints indeed worsen by a factor $\sim10$.
For larger values of $g$ non-linearity starts to have a more pronounced effect and constraints worsen even more.

This effect does not occur for the quadratic expansion.
The reason for that is that in a peak-background split, $\zeta$ transformed as~\cite{C1}
\begin{align}
\zeta=(1+2f\zeta_\mr l)\zeta_\mr s+f\bp{\zeta_\mr s^2-\sigma_s^2}
\end{align}
and $C_1$ as eq. \eqref{eq: quadratic}
\begin{align}
C_1=(1+2f\zeta_\mr l)C_\mr G+\widetilde{f}C_\mr G^2.
\end{align}
In both cases, the perturbation lies only in the linear term with identical prefactor $2f\zeta_\mr l$ and suppression by a factor $\gamma$ is absent here.
Therefore we expect that the constraints on $f_\mr{PBH}$ found in this work would be similar to those found by \emph{Young and Byrnes}.

\paragraph{Symmetrisation of constraints for cubic models}
To explain why the constraints for cubic expansions are more symmetric for positive and negative values of $g$, we instead need to consider that the abundance of PBHs was found by integration.
In this work, this integration was performed over $C_\mr G$, while in the work of \emph{Young and Byrnes} this was done over $\zeta_\mr s$.
The range of integration in this work was determined from $C_1=C_\mr{1,c}$ and $C_1=C_\mr{1,to}$, while that of \emph{Young and Byrnes} was determined from $\zeta=\zeta_\mr c$.
This is the critical value of $\zeta$ above which PBHs form, equivalent to the critical value of the density perturbation $\delta_\mr c$.
It was found in ref.~\cite{zetac1,zetac2}, that this takes the value $\zeta_\mr c=1$.

Because $\zeta_\mr l\ll1$, we can ignore all terms that depend on $\zeta_\mr l$ in order to find the solutions for $\zeta_\mr s$ and $C_\mr G$ of eqs. \eqref{eq: zeta cubic reiteration} and \eqref{eq: cubic reiteration} that correspond to these values.
We find that for a given $\zeta=\mc O(1)$ and $C_1=\mc O(1)$, the solution for $\zeta_\mr s$ and $C_\mr G$ are found respectively by good approximation as
\begin{align}
\zeta&\approx\zeta_\mr s+g\zeta_\mr s^3,\label{eq: zeta approx}\\
C_1&\approx C_\mr G+\widetilde{g}C_\mr G^3.\label{eq: C1 approx}
\end{align}
Although these equations have the same form, there is again suppression in the equation of $C_1$ since $\widetilde{g}$ is smaller than $g$ by a factor $3\gamma^2=\mc O(0.01)$.
Therefore, for $|g|\lesssim1$, the solution to $C_\mr G$ will be dominated by the linear term of eq. \eqref{eq: C1 approx}, with the cubic term offering only a slight perturbation.
This means that for a given magnitude $\bv g$, solutions for $C_\mr G$ for $g>0$ will be very close to those for $g<0$, which symmetrises the constraints.
The main difference between $g>0$ and $g<0$ in the constraints then comes from the different magnitude of the bounds of $b$ in eq. \eqref{eq: b constraints} for positive and negative values of $b$, or equivalently positive and negative values of $g$.

Meanwhile, the cubic term of eq. \eqref{eq: zeta approx} only depends on $g$.
Thus, for $|g|\lesssim1$ the cubic term could have a more appreciable effect on the solutions $C_\mr G$, and solutions for $g>0$ could deviate significantly from those for $g<0$, which asymmetrises the constraints.
Given that our work relies on perturbations in $C_1$, while that of \emph{Young and Byrnes} relies on perturbations in $\zeta$, we find that our constraints on cubic expansions is more symmetric between positive and negative values of $g$.

The same symmetrisation can actually also be observed in the quadratic model in figure \ref{fig: rPBH}, albeit with a lesser effect.
Solutions to $\zeta$ and $C_1$ for quadratic models can roughly be found by solving the equations
\begin{align}
\zeta&\approx\zeta_\mr s+f\zeta_\mr s^2,\\
C_1&\approx C_\mr G+\widetilde{f}C_\mr G^2.
\end{align}
Similar to before, $\widetilde{f}$ is smaller than $f$ by a factor $2\gamma=\mc O(0.1)$, which means that the quadratic term has a more appreciable effect on the solutions to $\zeta_\mr s$ than it has on $C_\mr G$, which makes our results more symmetrical compared to those of \emph{Young and Byrnes}.

\subsubsection{Constraints for higher order terms}
With the knowledge of the broadening of the cubic expansion we are now also in the position to understand what the effect of adding higher order terms to the non-Gaussian expansion of $C_1$ would be.
Following the patterns of eq. \eqref{eq: zeta transformation} and \eqref{eq: C1 ur} we find that $\zeta$ and $C_1$ transform for an $n$-th order term as
\begin{align}
\zeta&=\zeta_\mr G+h\bp{\zeta_\mr G^n-\ba{\zeta_\mr G^n}},\\
C_1&=C_\mr G\bp{1+nh\zeta_\mr G^{n-1}},
\end{align}
with $h$ the prefactor of the $n$-th order non-Gaussianity term and we ahve subtracted $\ba{\zeta_\mr G^n}$ to ensure that $\ba\zeta=0$.
Expanding $\zeta_\mr G=\zeta_\mr s+\zeta_\mr l$ in short and long wavelength modes, Taylor expanding the resulting equations for small $\zeta_l$ and ignoring again any terms that depend exclusively on $\zeta_\mr l$ gives for these two equations
\begin{align}
\zeta&=\zeta_\mr s+nh\zeta_\mr l\bp{\zeta_\mr s^{n-1}-\ba{\zeta_\mr G^{n-1}}}+h\bp{\zeta_\mr s^n-\ba{\zeta_\mr G^n}},\label{eq: zeta higher order term}\\
C_1&=C_\mr G+n(n-1)h\gamma^{n-2}\zeta_\mr lC_\mr G^{n-1}+h'C_\mr G^n,\label{eq: C1 higher order term}
\end{align}
where $h'=nh\gamma^{n-1}$.
In the first of these equations we have subtracted $\ba{\zeta_\mr G^{n-1}}$ to again ensure that $\ba{\zeta}=0$, in case $n$ is even.

The bias factor $b$ is determined by the leading order term in $\zeta_\mr l$.
Having ignored all higher order terms in $\zeta_\mr l$, we see that due to the factor $\gamma^{n-2}$, the second term in eq. \eqref{eq: C1 higher order term} has an increasingly less dominating effect for higher orders $n$.
As with the cubic term, the term $\gamma^{n-2}$ ensures that higher order terms will be subject to more intense broadening as larger values of $h$ are needed to produce the same value for $b$, leading in turn to weaker constraints on the non-Gaussianity parameter $h$.
This effect leads to particularly significant deviation from the constraints found using the approach of \emph{Young and Byrnes} as the second term in eq. \eqref{eq: zeta higher order term} is larger than that in eq. \eqref{eq: C1 higher order term} by an order $\gamma^{2-n}=\mc O(10^{n-2})$.

\section{Conclusion}
\label{sec: Conclusion}
PBHs offer very tight constraints on non-Gaussianity even if they make up only a small portion of the DM.
Because non-Gaussianity is an important prediction of many models of inflation, PBHs are a great window to study the early universe.

Previous work by \emph{Young and Byrnes}~\cite{C1} and \emph{Tada and Yokoyama}~\cite{bias} considered how the local-type non-Gaussianity parameters could be constrained from isocurvature modes appearing from fluctuations in the formation rate of PBHs, if PBHs make up a non-negligible fraction of dark matter.
This was done by assuming that PBHs form in regions where the curvature perturbation exceeds a critical value and using Press-Schechter theory to calculate the abundance of PBHs.
By considering modal coupling arising from non-Gaussianity, the peak-background split was applied to calculate the isocurvature modes appearing from fluctuations in the PBH formation rate.
By then using constraints from the \emph{Planck Collaboration}~\cite{Planck} on isocurvature modes they were able to constrain the non-Gaussianity parameters $f=3f_\mr{NL,local}/5$ and $g=9g_\mr{NL,local}/25$ for a given fraction of DM that is made up of PBHs, $f_\mr{PBH}$.

In this work, we have updated these constraints by using recent developments in the field, improving the calculation of the PBH abundance, and providing more accurate results.
These include the use of peaks theory instead of Press-Schechter theory, the use of the compaction instead of the curvature perturbation to describe when PBHs form and the correct mass scaling of the PBH mass (which depends on the scale and amplitude of the perturbation which formed the PBH).
Applying again non-Gaussianity when calculating the PBH abundance and applying the peak-background split to calculate the isocurvature modes, we found updated constraints on the non-Gaussianity parameters by using again the \emph{Planck} data~\cite{Planck}. Based on refs. \cite{C2,Ferrante:2022mui}, we consider only a narrow peak in the power spectrum to be responsible for PBH formation, ensuring the validity of the local-type expansion considered. The calculation could be extended to predict the isocurvature modes for specific models, such as the curvaton model considered in \cite{nonGaussPBHs3,Ferrante:2022mui}.

Similar to the results found by \emph{Young and Byrnes} and \emph{Tada and Yokoyama}, our updated calculation for quadratic models of non-Gaussianity offers tight constraints  of $\bv f\lesssim10^{-4}$ for $f_\mr{PBH}=1$, when all of the DM is made up of PBHs.
This means that the constraints earlier found by \emph{Young and Byrnes} for quadratic models through Press-Schechter theory still provide a reasonable approximation in case the easier approach of Press-Schechter theory is desired.

For cubic models of non-Gaussianity the constraints in this work weaken significantly compared to those found by \emph{Young and Byrnes}, however.
This is due to additional suppression of cubic order non-Gaussian terms by a factor $\gamma\sim0.1$.
For smaller values of $g$ this broadens constraints by a factor $\mc O(10)$, while for larger values of $g$ the equations start to become non-linear, and constraints worsen even more.
Even so, for large values of $f_\mr{PBH}$, even cubic models of non-Gaussianity can still be tightly constrained by the updated calculation, with $\bv g\lesssim10^{-3}$ for $f_\mr{PBH}=1$.

We emphasise that the constraints calculated in this paper assume that the local model of non-Gaussianity is accurate over a large range of scales; from the large scales visible in the CMB to the small scales at which PBHs form.
The constraints therefore apply specifically to the local-type non-Gaussianity parameters.
A local-type bispectrum peaks in the squeezed-limit, implying a strong correlation between large and small scales --- giving strong constraints on the non-Gaussianity parameter. However, other bispectrum shapes (such as equilateral and orthogonal) do not peak in this limit, implying a weak(er) correlation --- which would result in much weaker constraints.

In the case of models predicting varying levels of non-Gaussianity on different scales, the constraints calculated here would not be accurate. For example, if the curvaton model is considered, then the small-scale perturbations can be sourced by the curvaton, which can be decoupled from the inflaton, which sources the large-scale perturbations. This would imply a much weaker correlation between the small and large scales, and correspondingly weaker constraints on the non-Gaussianity parameters. Such constraints could be computed on a model-by-model basis using the methods presented here, although such consideration is beyond the scope of this paper. 
However, the constraints on the non-Gaussianity parameters from PBH isocurvature modes could still be competitive with those available from other observations (such as the CMB) even if weaker by several orders of magnitude.

We also briefly considered the constraints on higher-order non-Gaussianity parameters. \emph{Young and Byrnes} previously found that the isocurvature constraints applied (almost) equally at all orders of non-Gaussianity and would essentially rule out non-Gaussianity at all orders if PBHs are detected. However, we have demonstrated here that this is not the case, and that the constraints become significantly weaker as higher order terms are considered.

\acknowledgments

RvL is supported by the Swiss National Science Foundation grant No 207739. SY is an MCSA postdoctoral fellow, and this project has received funding from the European Union’s Horizon 2020 research and innovation programme under the Marie Skłodowska-Curie grant agreement No 101029832. The authors thank Subodh Patil and Alessandra Silvestri for useful feedback.

\bibliographystyle{JHEP}
\bibliography{main}

\providecommand{\href}[2]{#2}\begingroup\raggedright\begin{thebibliography}{10}

\bibitem{Clesse:2017bsw}
S.~Clesse and J.~Garc\'\i{}a-Bellido, \emph{{Seven Hints for Primordial Black
  Hole Dark Matter}},
  \href{https://doi.org/10.1016/j.dark.2018.08.004}{\emph{Phys. Dark Univ.}
  {\bfseries 22} (2018) 137}
  [\href{https://arxiv.org/abs/1711.10458}{{\ttfamily 1711.10458}}].

\bibitem{DMoverview}
B.~Carr and F.~Kuhnel, \emph{{Primordial Black Holes as Dark Matter: Recent
  Developments}},
  \href{https://doi.org/10.1146/annurev-nucl-050520-125911}{\emph{Ann. Rev.
  Nucl. Part. Sci.} {\bfseries 70} (2020) 355}
  [\href{https://arxiv.org/abs/2006.02838}{{\ttfamily 2006.02838}}].

\bibitem{SMBHs}
R.~Bean and J.~Magueijo, \emph{{Could supermassive black holes be
  quintessential primordial black holes?}},
  \href{https://doi.org/10.1103/PhysRevD.66.063505}{\emph{Phys. Rev. D}
  {\bfseries 66} (2002) 063505}
  [\href{https://arxiv.org/abs/astro-ph/0204486}{{\ttfamily
  astro-ph/0204486}}].

\bibitem{LIGOVIRGO}
S.~Bird, I.~Cholis, J.B.~Mu\~noz, Y.~Ali-Ha\"\i{}moud, M.~Kamionkowski,
  E.D.~Kovetz et~al., \emph{{Did LIGO detect dark matter?}},
  \href{https://doi.org/10.1103/PhysRevLett.116.201301}{\emph{Phys. Rev. Lett.}
  {\bfseries 116} (2016) 201301}
  [\href{https://arxiv.org/abs/1603.00464}{{\ttfamily 1603.00464}}].

\bibitem{strings1}
S.W.~Hawking, \emph{{Black Holes From Cosmic Strings}},
  \href{https://doi.org/10.1016/0370-2693(89)90206-2}{\emph{Phys. Lett. B}
  {\bfseries 231} (1989) 237}.

\bibitem{strings2}
A.~Polnarev and R.~Zembowicz, \emph{{Formation of Primordial Black Holes by
  Cosmic Strings}}, \href{https://doi.org/10.1103/PhysRevD.43.1106}{\emph{Phys.
  Rev. D} {\bfseries 43} (1991) 1106}.

\bibitem{strings3}
R.N.~Hansen, M.~Christensen and A.L.~Larsen, \emph{{Cosmic string loops
  collapsing to black holes}},
  \href{https://doi.org/10.1142/S0217751X00001450}{\emph{Int. J. Mod. Phys. A}
  {\bfseries 15} (2000) 4433}
  [\href{https://arxiv.org/abs/gr-qc/9902048}{{\ttfamily gr-qc/9902048}}].

\bibitem{strings4}
C.J.~Hogan, \emph{{MASSIVE BLACK HOLES GENERATED BY COSMIC STRINGS}},
  \href{https://doi.org/10.1016/0370-2693(84)90810-4}{\emph{Phys. Lett. B}
  {\bfseries 143} (1984) 87}.

\bibitem{strings5}
M.~Nagasawa, \emph{{Primordial black hole formation by stabilized embedded
  strings in the early universe}},
  \href{https://doi.org/10.1007/s10714-005-0141-9}{\emph{Gen. Rel. Grav.}
  {\bfseries 37} (2005) 1635}.

\bibitem{strings6}
C.~James-Turner, D.P.B.~Weil, A.M.~Green and E.J.~Copeland, \emph{{Constraints
  on the cosmic string loop collapse fraction from primordial black holes}},
  \href{https://doi.org/10.1103/PhysRevD.101.123526}{\emph{Phys. Rev. D}
  {\bfseries 101} (2020) 123526}
  [\href{https://arxiv.org/abs/1911.12658}{{\ttfamily 1911.12658}}].

\bibitem{bubble1}
M.~Crawford and D.N.~Schramm, \emph{{Spontaneous Generation of Density
  Perturbations in the Early Universe}},
  \href{https://doi.org/10.1038/298538a0}{\emph{Nature} {\bfseries 298} (1982)
  538}.

\bibitem{bubble2}
S.W.~Hawking, I.G.~Moss and J.M.~Stewart, \emph{{Bubble Collisions in the Very
  Early Universe}}, \href{https://doi.org/10.1103/PhysRevD.26.2681}{\emph{Phys.
  Rev. D} {\bfseries 26} (1982) 2681}.

\bibitem{bubble3}
D.~La and P.J.~Steinhardt, \emph{{Bubble Percolation in Extended Inflationary
  Models}}, \href{https://doi.org/10.1016/0370-2693(89)90890-3}{\emph{Phys.
  Lett. B} {\bfseries 220} (1989) 375}.

\bibitem{bubble4}
I.G.~Moss, \emph{{Singularity formation from colliding bubbles}},
  \href{https://doi.org/10.1103/PhysRevD.50.676}{\emph{Phys. Rev. D} {\bfseries
  50} (1994) 676}.

\bibitem{bubble5}
H.~Kodama, M.~Sasaki and K.~Sato, \emph{{Abundance of Primordial Holes Produced
  by Cosmological First Order Phase Transition}},
  \href{https://doi.org/10.1143/PTP.68.1979}{\emph{Prog. Theor. Phys.}
  {\bfseries 68} (1982) 1979}.

\bibitem{scalar}
E.~Cotner and A.~Kusenko, \emph{{Primordial black holes from scalar field
  evolution in the early universe}},
  \href{https://doi.org/10.1103/PhysRevD.96.103002}{\emph{Phys. Rev. D}
  {\bfseries 96} (2017) 103002}
  [\href{https://arxiv.org/abs/1706.09003}{{\ttfamily 1706.09003}}].

\bibitem{wall1}
S.G.~Rubin, M.Y.~Khlopov and A.S.~Sakharov, \emph{{Primordial black holes from
  nonequilibrium second order phase transition}}, {\emph{Grav. Cosmol.}
  {\bfseries 6} (2000) 51}
  [\href{https://arxiv.org/abs/hep-ph/0005271}{{\ttfamily hep-ph/0005271}}].

\bibitem{wall2}
S.G.~Rubin, A.S.~Sakharov and M.Y.~Khlopov, \emph{{The Formation of primary
  galactic nuclei during phase transitions in the early universe}},
  \href{https://doi.org/10.1134/1.1385631}{\emph{J. Exp. Theor. Phys.}
  {\bfseries 91} (2001) 921}
  [\href{https://arxiv.org/abs/hep-ph/0106187}{{\ttfamily hep-ph/0106187}}].

\bibitem{wall3}
V.~Dokuchaev, Y.~Eroshenko and S.~Rubin, \emph{{Quasars formation around
  clusters of primordial black holes}}, {\emph{Grav. Cosmol.} {\bfseries 11}
  (2005) 99} [\href{https://arxiv.org/abs/astro-ph/0412418}{{\ttfamily
  astro-ph/0412418}}].

\bibitem{singl1}
C.~Pattison, V.~Vennin, H.~Assadullahi and D.~Wands, \emph{{Quantum diffusion
  during inflation and primordial black holes}},
  \href{https://doi.org/10.1088/1475-7516/2017/10/046}{\emph{JCAP} {\bfseries
  10} (2017) 046} [\href{https://arxiv.org/abs/1707.00537}{{\ttfamily
  1707.00537}}].

\bibitem{singl2}
J.M.~Ezquiaga and J.~Garc\'\i{}a-Bellido, \emph{{Quantum diffusion beyond
  slow-roll: implications for primordial black-hole production}},
  \href{https://doi.org/10.1088/1475-7516/2018/08/018}{\emph{JCAP} {\bfseries
  08} (2018) 018} [\href{https://arxiv.org/abs/1805.06731}{{\ttfamily
  1805.06731}}].

\bibitem{singl3}
M.~Biagetti, G.~Franciolini, A.~Kehagias and A.~Riotto, \emph{{Primordial Black
  Holes from Inflation and Quantum Diffusion}},
  \href{https://doi.org/10.1088/1475-7516/2018/07/032}{\emph{JCAP} {\bfseries
  07} (2018) 032} [\href{https://arxiv.org/abs/1804.07124}{{\ttfamily
  1804.07124}}].

\bibitem{singl4}
A.M.~Green and K.A.~Malik, \emph{{Primordial black hole production due to
  preheating}}, \href{https://doi.org/10.1103/PhysRevD.64.021301}{\emph{Phys.
  Rev. D} {\bfseries 64} (2001) 021301}
  [\href{https://arxiv.org/abs/hep-ph/0008113}{{\ttfamily hep-ph/0008113}}].

\bibitem{singl5}
B.A.~Bassett and S.~Tsujikawa, \emph{{Inflationary preheating and primordial
  black holes}}, \href{https://doi.org/10.1103/PhysRevD.63.123503}{\emph{Phys.
  Rev. D} {\bfseries 63} (2001) 123503}
  [\href{https://arxiv.org/abs/hep-ph/0008328}{{\ttfamily hep-ph/0008328}}].

\bibitem{Kawai:2021edk}
S.~Kawai and J.~Kim, \emph{{Primordial black holes from Gauss-Bonnet-corrected
  single field inflation}},
  \href{https://doi.org/10.1103/PhysRevD.104.083545}{\emph{Phys. Rev. D}
  {\bfseries 104} (2021) 083545}
  [\href{https://arxiv.org/abs/2108.01340}{{\ttfamily 2108.01340}}].

\bibitem{mingfl1}
L.~Randall, M.~Soljacic and A.H.~Guth, \emph{{Supernatural inflation: Inflation
  from supersymmetry with no (very) small parameters}},
  \href{https://doi.org/10.1016/0550-3213(96)00174-5}{\emph{Nucl. Phys. B}
  {\bfseries 472} (1996) 377}
  [\href{https://arxiv.org/abs/hep-ph/9512439}{{\ttfamily hep-ph/9512439}}].

\bibitem{mingfl2}
J.~Garcia-Bellido, A.D.~Linde and D.~Wands, \emph{{Density perturbations and
  black hole formation in hybrid inflation}},
  \href{https://doi.org/10.1103/PhysRevD.54.6040}{\emph{Phys. Rev. D}
  {\bfseries 54} (1996) 6040}
  [\href{https://arxiv.org/abs/astro-ph/9605094}{{\ttfamily
  astro-ph/9605094}}].

\bibitem{Kawai:2022emp}
S.~Kawai and J.~Kim, \emph{{Primordial black holes and gravitational waves from
  nonminimally coupled supergravity inflation}},
  \href{https://doi.org/10.1103/PhysRevD.107.043523}{\emph{Phys. Rev. D}
  {\bfseries 107} (2023) 043523}
  [\href{https://arxiv.org/abs/2209.15343}{{\ttfamily 2209.15343}}].

\bibitem{densitycollapse}
B.J.~Carr and S.W.~Hawking, \emph{{Black holes in the early Universe}},
  {\emph{Mon. Not. Roy. Astron. Soc.} {\bfseries 168} (1974) 399}.

\bibitem{Carr3}
B.J.~Carr, \emph{{The Primordial black hole mass spectrum}},
  \href{https://doi.org/10.1086/153853}{\emph{Astrophys. J.} {\bfseries 201}
  (1975) 1}.

\bibitem{deltac1}
J.C.~Niemeyer and K.~Jedamzik, \emph{{Dynamics of primordial black hole
  formation}}, \href{https://doi.org/10.1103/PhysRevD.59.124013}{\emph{Phys.
  Rev. D} {\bfseries 59} (1999) 124013}
  [\href{https://arxiv.org/abs/astro-ph/9901292}{{\ttfamily
  astro-ph/9901292}}].

\bibitem{deltac2}
I.~Hawke and J.M.~Stewart, \emph{{The dynamics of primordial black hole
  formation}}, \href{https://doi.org/10.1088/0264-9381/19/14/310}{\emph{Class.
  Quant. Grav.} {\bfseries 19} (2002) 3687}.

\bibitem{deltac3}
I.~Musco, J.C.~Miller and L.~Rezzolla, \emph{{Computations of primordial black
  hole formation}},
  \href{https://doi.org/10.1088/0264-9381/22/7/013}{\emph{Class. Quant. Grav.}
  {\bfseries 22} (2005) 1405}
  [\href{https://arxiv.org/abs/gr-qc/0412063}{{\ttfamily gr-qc/0412063}}].

\bibitem{deltac4}
I.~Musco, J.C.~Miller and A.G.~Polnarev, \emph{{Primordial black hole formation
  in the radiative era: Investigation of the critical nature of the collapse}},
  \href{https://doi.org/10.1088/0264-9381/26/23/235001}{\emph{Class. Quant.
  Grav.} {\bfseries 26} (2009) 235001}
  [\href{https://arxiv.org/abs/0811.1452}{{\ttfamily 0811.1452}}].

\bibitem{deltac5}
T.~Harada, C.-M.~Yoo and K.~Kohri, \emph{{Threshold of primordial black hole
  formation}}, \href{https://doi.org/10.1103/PhysRevD.88.084051}{\emph{Phys.
  Rev. D} {\bfseries 88} (2013) 084051}
  [\href{https://arxiv.org/abs/1309.4201}{{\ttfamily 1309.4201}}].

\bibitem{deltac6}
T.~Nakama, T.~Harada, A.G.~Polnarev and J.~Yokoyama, \emph{{Identifying the
  most crucial parameters of the initial curvature profile for primordial black
  hole formation}},
  \href{https://doi.org/10.1088/1475-7516/2014/01/037}{\emph{JCAP} {\bfseries
  01} (2014) 037} [\href{https://arxiv.org/abs/1310.3007}{{\ttfamily
  1310.3007}}].

\bibitem{YoungMuscoByrnes}
S.~Young, I.~Musco and C.T.~Byrnes, \emph{{Primordial black hole formation and
  abundance: contribution from the non-linear relation between the density and
  curvature perturbation}},
  \href{https://doi.org/10.1088/1475-7516/2019/11/012}{\emph{JCAP} {\bfseries
  11} (2019) 012} [\href{https://arxiv.org/abs/1904.00984}{{\ttfamily
  1904.00984}}].

\bibitem{Niemeyer1}
J.C.~Niemeyer and K.~Jedamzik, \emph{{Near-critical gravitational collapse and
  the initial mass function of primordial black holes}},
  \href{https://doi.org/10.1103/PhysRevLett.80.5481}{\emph{Phys. Rev. Lett.}
  {\bfseries 80} (1998) 5481}
  [\href{https://arxiv.org/abs/astro-ph/9709072}{{\ttfamily
  astro-ph/9709072}}].

\bibitem{Carr}
B.~Carr, K.~Kohri, Y.~Sendouda and J.~Yokoyama, \emph{{Constraints on
  primordial black holes}},
  \href{https://doi.org/10.1088/1361-6633/ac1e31}{\emph{Rept. Prog. Phys.}
  {\bfseries 84} (2021) 116902}
  [\href{https://arxiv.org/abs/2002.12778}{{\ttfamily 2002.12778}}].

\bibitem{nonGaussPBHs1}
J.S.~Bullock and J.R.~Primack, \emph{{NonGaussian fluctuations and primordial
  black holes from inflation}},
  \href{https://doi.org/10.1103/PhysRevD.55.7423}{\emph{Phys. Rev. D}
  {\bfseries 55} (1997) 7423}
  [\href{https://arxiv.org/abs/astro-ph/9611106}{{\ttfamily
  astro-ph/9611106}}].

\bibitem{nonGaussPBHs2}
C.T.~Byrnes, E.J.~Copeland and A.M.~Green, \emph{{Primordial black holes as a
  tool for constraining non-Gaussianity}},
  \href{https://doi.org/10.1103/PhysRevD.86.043512}{\emph{Phys. Rev. D}
  {\bfseries 86} (2012) 043512}
  [\href{https://arxiv.org/abs/1206.4188}{{\ttfamily 1206.4188}}].

\bibitem{nonGaussPBHs3}
S.~Young and C.T.~Byrnes, \emph{{Primordial black holes in non-Gaussian
  regimes}}, \href{https://doi.org/10.1088/1475-7516/2013/08/052}{\emph{JCAP}
  {\bfseries 08} (2013) 052} [\href{https://arxiv.org/abs/1307.4995}{{\ttfamily
  1307.4995}}].

\bibitem{Ferrante:2022mui}
G.~Ferrante, G.~Franciolini, A.~Iovino, Junior. and A.~Urbano,
  \emph{{Primordial non-gaussianity up to all orders: theoretical aspects and
  implications for primordial black hole models}},
  \href{https://arxiv.org/abs/2211.01728}{{\ttfamily 2211.01728}}.

\bibitem{Matsubara:2022nbr}
T.~Matsubara and M.~Sasaki, \emph{{Non-Gaussianity effects on the primordial
  black hole abundance for sharply-peaked primordial spectrum}},
  \href{https://doi.org/10.1088/1475-7516/2022/10/094}{\emph{JCAP} {\bfseries
  10} (2022) 094} [\href{https://arxiv.org/abs/2208.02941}{{\ttfamily
  2208.02941}}].

\bibitem{C1}
S.~Young and C.T.~Byrnes, \emph{{Signatures of non-gaussianity in the
  isocurvature modes of primordial black hole dark matter}},
  \href{https://doi.org/10.1088/1475-7516/2015/04/034}{\emph{JCAP} {\bfseries
  04} (2015) 034} [\href{https://arxiv.org/abs/1503.01505}{{\ttfamily
  1503.01505}}].

\bibitem{Planck}
{\scshape Planck} collaboration, \emph{{Planck 2015 results. XX. Constraints on
  inflation}}, \href{https://doi.org/10.1051/0004-6361/201525898}{\emph{Astron.
  Astrophys.} {\bfseries 594} (2016) A20}
  [\href{https://arxiv.org/abs/1502.02114}{{\ttfamily 1502.02114}}].

\bibitem{C2}
S.~Young, \emph{{Peaks and primordial black holes: the~effect of
  non-Gaussianity}},
  \href{https://doi.org/10.1088/1475-7516/2022/05/037}{\emph{JCAP} {\bfseries
  05} (2022) 037} [\href{https://arxiv.org/abs/2201.13345}{{\ttfamily
  2201.13345}}].

\bibitem{WMAP}
{\scshape WMAP} collaboration, \emph{{First year Wilkinson Microwave Anisotropy
  Probe (WMAP) observations: tests of gaussianity}},
  \href{https://doi.org/10.1086/377220}{\emph{Astrophys. J. Suppl.} {\bfseries
  148} (2003) 119} [\href{https://arxiv.org/abs/astro-ph/0302223}{{\ttfamily
  astro-ph/0302223}}].

\bibitem{nonGauss2}
N.~Bartolo, E.~Komatsu, S.~Matarrese and A.~Riotto, \emph{{Non-Gaussianity from
  inflation: Theory and observations}},
  \href{https://doi.org/10.1016/j.physrep.2004.08.022}{\emph{Phys. Rept.}
  {\bfseries 402} (2004) 103}
  [\href{https://arxiv.org/abs/astro-ph/0406398}{{\ttfamily
  astro-ph/0406398}}].

\bibitem{OmegaPBH}
C.T.~Byrnes, M.~Hindmarsh, S.~Young and M.R.S.~Hawkins, \emph{{Primordial black
  holes with an accurate QCD equation of state}},
  \href{https://doi.org/10.1088/1475-7516/2018/08/041}{\emph{JCAP} {\bfseries
  08} (2018) 041} [\href{https://arxiv.org/abs/1801.06138}{{\ttfamily
  1801.06138}}].

\bibitem{Meq}
T.~Nakama, J.~Silk and M.~Kamionkowski, \emph{{Stochastic gravitational waves
  associated with the formation of primordial black holes}},
  \href{https://doi.org/10.1103/PhysRevD.95.043511}{\emph{Phys. Rev. D}
  {\bfseries 95} (2017) 043511}
  [\href{https://arxiv.org/abs/1612.06264}{{\ttfamily 1612.06264}}].

\bibitem{OmegaCDM}
{\scshape Planck} collaboration, \emph{{Planck 2018 results. X. Constraints on
  inflation}}, \href{https://doi.org/10.1051/0004-6361/201833887}{\emph{Astron.
  Astrophys.} {\bfseries 641} (2020) A10}
  [\href{https://arxiv.org/abs/1807.06211}{{\ttfamily 1807.06211}}].

\bibitem{bias}
Y.~Tada and S.~Yokoyama, \emph{{Primordial black holes as biased tracers}},
  \href{https://doi.org/10.1103/PhysRevD.91.123534}{\emph{Phys. Rev. D}
  {\bfseries 91} (2015) 123534}
  [\href{https://arxiv.org/abs/1502.01124}{{\ttfamily 1502.01124}}].

\bibitem{pk}
J.M.~Bardeen, J.R.~Bond, N.~Kaiser and A.S.~Szalay, \emph{{The Statistics of
  Peaks of Gaussian Random Fields}},
  \href{https://doi.org/10.1086/164143}{\emph{Astrophys. J.} {\bfseries 304}
  (1986) 15}.

\bibitem{pk1}
S.~Young and M.~Musso, \emph{{Application of peaks theory to the abundance of
  primordial black holes}},
  \href{https://doi.org/10.1088/1475-7516/2020/11/022}{\emph{JCAP} {\bfseries
  11} (2020) 022} [\href{https://arxiv.org/abs/2001.06469}{{\ttfamily
  2001.06469}}].

\bibitem{pk2}
C.~Germani and R.K.~Sheth, \emph{{Nonlinear statistics of primordial black
  holes from Gaussian curvature perturbations}},
  \href{https://doi.org/10.1103/PhysRevD.101.063520}{\emph{Phys. Rev. D}
  {\bfseries 101} (2020) 063520}
  [\href{https://arxiv.org/abs/1912.07072}{{\ttfamily 1912.07072}}].

\bibitem{Compaction1}
I.~Musco, \emph{{Threshold for primordial black holes: Dependence on the shape
  of the cosmological perturbations}},
  \href{https://doi.org/10.1103/PhysRevD.100.123524}{\emph{Phys. Rev. D}
  {\bfseries 100} (2019) 123524}
  [\href{https://arxiv.org/abs/1809.02127}{{\ttfamily 1809.02127}}].

\bibitem{Compaction2}
S.~Young, \emph{{The primordial black hole formation criterion re-examined:
  Parametrisation, timing and the choice of window function}},
  \href{https://doi.org/10.1142/S0218271820300025}{\emph{Int. J. Mod. Phys. D}
  {\bfseries 29} (2019) 2030002}
  [\href{https://arxiv.org/abs/1905.01230}{{\ttfamily 1905.01230}}].

\bibitem{TypeII}
M.~Kopp, S.~Hofmann and J.~Weller, \emph{{Separate Universes Do Not Constrain
  Primordial Black Hole Formation}},
  \href{https://doi.org/10.1103/PhysRevD.83.124025}{\emph{Phys. Rev. D}
  {\bfseries 83} (2011) 124025}
  [\href{https://arxiv.org/abs/1012.4369}{{\ttfamily 1012.4369}}].

\bibitem{Gow:2020bzo}
A.D.~Gow, C.T.~Byrnes, P.S.~Cole and S.~Young, \emph{{The power spectrum on
  small scales: Robust constraints and comparing PBH methodologies}},
  \href{https://doi.org/10.1088/1475-7516/2021/02/002}{\emph{JCAP} {\bfseries
  02} (2021) 002} [\href{https://arxiv.org/abs/2008.03289}{{\ttfamily
  2008.03289}}].

\bibitem{PBHparameters}
N.~Kitajima, Y.~Tada, S.~Yokoyama and C.-M.~Yoo, \emph{{Primordial black holes
  in peak theory with a non-Gaussian tail}},
  \href{https://doi.org/10.1088/1475-7516/2021/10/053}{\emph{JCAP} {\bfseries
  10} (2021) 053} [\href{https://arxiv.org/abs/2109.00791}{{\ttfamily
  2109.00791}}].

\bibitem{zetac2}
S.~Young, C.T.~Byrnes and M.~Sasaki, \emph{{Calculating the mass fraction of
  primordial black holes}},
  \href{https://doi.org/10.1088/1475-7516/2014/07/045}{\emph{JCAP} {\bfseries
  07} (2014) 045} [\href{https://arxiv.org/abs/1405.7023}{{\ttfamily
  1405.7023}}].

\bibitem{zetac1}
M.~Shibata and M.~Sasaki, \emph{{Black hole formation in the Friedmann
  universe: Formulation and computation in numerical relativity}},
  \href{https://doi.org/10.1103/PhysRevD.60.084002}{\emph{Phys. Rev. D}
  {\bfseries 60} (1999) 084002}
  [\href{https://arxiv.org/abs/gr-qc/9905064}{{\ttfamily gr-qc/9905064}}].

\end{thebibliography}\endgroup

\end{document}